\documentclass[11pt]{article}
\usepackage{geometry}                
\usepackage{graphicx}
\usepackage{amssymb,amsfonts,amsmath,url}
\usepackage{fullpage}
\usepackage[doublespacing]{setspace}
\usepackage{float} 
\usepackage{ifthen}
\usepackage{cell}
\usepackage{authblk}
\usepackage{subfigure}
\usepackage{caption}
\usepackage{booktabs}
\usepackage{color}
\begin{document}

\title{Historical Contingency and Entrenchment in Protein Evolution Under Purifying 
Selection}

\author[1,*]{Premal Shah}
\author[1]{David M.~McCandlish}
\author[1]{Joshua B.~Plotkin}
\affil[1]{University of Pennsylvania, Philadelphia PA 19104, USA}
\affil[*]{To whom correspondence should be addressed. Email: shahpr@sas.upenn.edu}
\date{}    

\maketitle

\abstract The fitness contribution of an allele at one genetic site may depend on alleles at other sites, a phenomenon known as epistasis. Epistasis can profoundly influence the process of evolution in populations under selection, and can shape the course of protein evolution across divergent species. Whereas epistasis between adaptive substitutions has been the subject of  extensive study, relatively little is known about epistasis under purifying selection. Here we use mechanistic models of thermodynamic stability in a ligand-binding protein to explore  the structure of epistatic interactions between substitutions that fix in protein sequences under purifying selection. We find that the selection coefficients of mutations that are nearly-neutral when they fix are highly contingent on the presence of preceding mutations. Conversely, mutations that are nearly-neutral when they fix are subsequently entrenched due to epistasis with later substitutions. Our evolutionary model includes insertions and deletions, as well as point mutations, and so it allows us to quantify epistasis within each of these classes of mutations, and also to study the evolution of protein length. We find that protein length remains largely constant over time, because indels are more deleterious than point mutations. Our results imply that, even under purifying selection, protein sequence evolution is highly contingent on history and so it cannot be predicted by the phenotypic effects of mutations assayed in the wild-type sequence.

\clearpage \pagebreak

\section{Introduction}
Whether any particular heritable mutation is advantageous or deleterious to an
organism often depends on the evolutionary history of the population. A mutation
that is beneficial at the time of its introduction may confer its beneficial
effect only in presence of other potentiating or permissive mutations that have
previously fixed in the population \cite{Weinreich:2005vg,Weinreich:2006ig,Bridgham:2009jv,Bloom:2010eu,McLaughlin:2012hw,Natarajan:2013gi}. 
Thus, the fate of a mutation arising in a
population is often {\em contingent} on previous mutations~\cite{JayGould:wh,beatty2006replaying,Travisano:1995uf}. Similarly, once a
mutation has fixed in a population, that focal mutation becomes part of the genetic
background onto which subsequent modifications are introduced.  Because the
beneficial effects of the subsequent modifications may depend on the focal
mutation, as time passes reversion of the focal mutation may become increasingly
deleterious, leading to a type of evolutionary conservatism, or {\em
entrenchment} \cite{Pollock:2012ge,Muller:1939if,Riedl:1977th,Wimsatt:1986ek,Raff:1996tv,szathmary2006path}. 

In the context of protein evolution, this situation is seen most clearly by
considering evolution as a sequence of single amino-acid changes \cite{Smith:1970uz} that extends both
forwards and backwards in time from some focal substitution. To assess the roles of
contingency and entrenchment we can study the degree to which the focal
substitution was facilitated by previous substitutions, and the degree to which
the focal substitution influences the subsequent course of evolution (Figure
\ref{f_concept}A).

Dependencies within a sequence of substitutions are closely connected to the
concept of epistasis -- that is, the idea that the fitness effect of a mutation at
a particular genetic site may depend on the genetic background in which it arises \cite{Whitlock:1995vj,Wolf:2000vt,Cordell:2002de,Phillips:2008dm}.
In the absence of epistasis, a mutation has the same effect regardless of its
context and therefore regardless of any prior or subsequent evolutionary history.
By contrast, in the presence of epistasis, each substitution may be contingent on
the entire prior history of the protein, and it may constrain all subsequent
evolution.

The potential for epistasis to play an important role in evolution, including
protein evolution, has certainly not been over-looked by researchers \cite{Kondrashov:2002cb,Depristo:2005jc,Weinreich:2006ig,Lunzer:2010co,Burke:2010eq,Kryazhimskiy:2011do,Khan:2011ik,Chou:2011dk,Breen:2012fd,Blount:2012jl,Gong:2013ju,Lang:2013fh,Wiser:2013ch};
nor have the concepts of contingency \cite{Travisano:1995uf,Losos:1998vg,Blount:2008fo,Bridgham:2009jv,Salverda:2011fa,Meyer:2012ef,Dickinson:2013hv,Harms:2014gv} and, more recently, entrenchment
\cite{Pollock:2012ge,Naumenko:2012bt,Soylemez:2012dq,Ashenberg:2013gd}.  However, most
studies have addressed the relations
between epistasis, contingency and entrenchment in the context of adaptive
evolution \cite{Weinreich:2006ig,Blount:2008fo,Bridgham:2009jv,Bloom:2010eu,Lunzer:2010co,Salverda:2011fa,Khan:2011ik,Chou:2011dk,McLaughlin:2012hw,Natarajan:2013gi,Dickinson:2013hv,Harms:2014gv}, whereas the consequences of epistasis under purifying
selection have received less attention~\cite{Bershtein:2006fu,Wang:2007bz,Povolotskaya:2010bc,Pollock:2012ge,Xu:2014gw}.  Indeed, although some more sophisticated models have been proposed, for e.g.~\cite{Pollock:1999bk,Robinson:2003cm,Rodrigue:2006ir,Bastolla:2006ck,Choi:2007ds,Arenas:2013cp}, all commonly used models of
long-term protein evolution assume that epistasis is absent so that all sites
evolve independently \cite{Goldman:1994wf,Kosiol:2007gf,Yang:2008dd,Rodrigue:2010by,Tamuri:2012hv}.

Here we explore the relationship between epistasis, contingency and entrenchment
under long-term purifying selection on protein stability.  Our analysis combines
computational models for protein structures with population-genetic models for
evolutionary dynamics.  We use homology models \cite{MartiRenom:2000ea} to characterize the effects of
point-mutations, insertions and deletions on a protein's stability and fitness.
These allow us to simulate evolutionary trajectories of protein sequences under
purifying selection, by sequential fixation of nearly-neutral mutations.  We then
dissect the epistatic relationships between these substitutions by systematically
inserting or reverting particular substitutions at various time-points along the
evolutionary trajectory.  

Using this framework, we demonstrate that even nearly-neutral mutations that 
fix under purifying selection tend to be highly epistatic with each other.  In
particular, we find that each substitution that fixes is typically permitted to
fix only because of the presence of preceding substitutions---that is, each
substitution would be too deleterious to fix were it not for epistasis with
preceding substitutions. Moreover, we find that substitutions typically become
entrenched over time by epistasis---so that a substitution, which was
nearly-neutral when it fixed, becomes increasingly deleterious to revert as
subsequent substitutions accumulate \cite{Pollock:2012ge,Naumenko:2012bt}. Finally, we show that the
pattern of epistasis along the evolutionary trajectory under nearly-neutral
evolution is very different from the pattern observed under adaptive evolution.
These results imply that protein evolution under purifying selection is highly
contingent on history, and that the long-term trajectory cannot be predicted by
assaying the effects of substitutions in an ancestral protein sequence.

\section{Methods}
\subsection{Evolutionary model}
We explore the evolution of a protein sequence in the regime of weak-mutation, so
that each new mutation introduced into the population either is lost or goes to
fixation, with probabilities that depend upon the mutant's fitness, before another
mutation is introduced.  Fixation or loss are considered instantaneous so that the
population is always monomorphic for a particular protein sequence.  We study the
238 amino-acid lysine-arginine-ornithine-binding periplasmic protein (\emph{argT})
from \emph{S.~typhimurium} as a model system, chosen because its crystal 
structure
is known and is simple enough that computational predictions for the effects of
mutations are feasible and credible.  

For each proposed mutant sequence we construct a homology model using
\texttt{Modeller} and we compute its Discrete Optimized Protein Energy (DOPE
score) \cite{Shen:2006dl,Eramian:2006eg,Eswar:2008ge} as a surrogate for its
thermodynamic stability. Empirical studies have shown that the DOPE
score is highly correlated with a protein's thermodynamic stability
\cite{Howell:2013vt}, and is widely used in studies of thermal
adaptation \cite{Howell:2013vt}, of ligand binding affinities
\cite{Genheden:2012bq}, as well as phylogenetic models of protein evolution
\cite{Arenas:2013cp}.  Alternative computational techniques for studying the
effects of mutations on stability include idealized lattice models
\cite{Bloom:2006ch}, simple contact potentials
\cite{Tokuriki:2007ij,Pollock:2012ge}, and even physical force-field models
\cite{Serohijos:2014ie}.  Compared to lattice models or contact potentials, the
homology models we use are computationally expensive, which restricts the number of mutants
that we can study, but they provide a more realistic description of mutational
effects on stability. Nonetheless, homology models cannot reliably predict large
changes in protein structures due to individual mutations, such as those studied by
very expensive, atomic-resolution molecular dynamic simulations (see
\cite{Dror:2012cs} and references within).

\begin{figure}[ht]
        \includegraphics[width=0.6\textwidth]{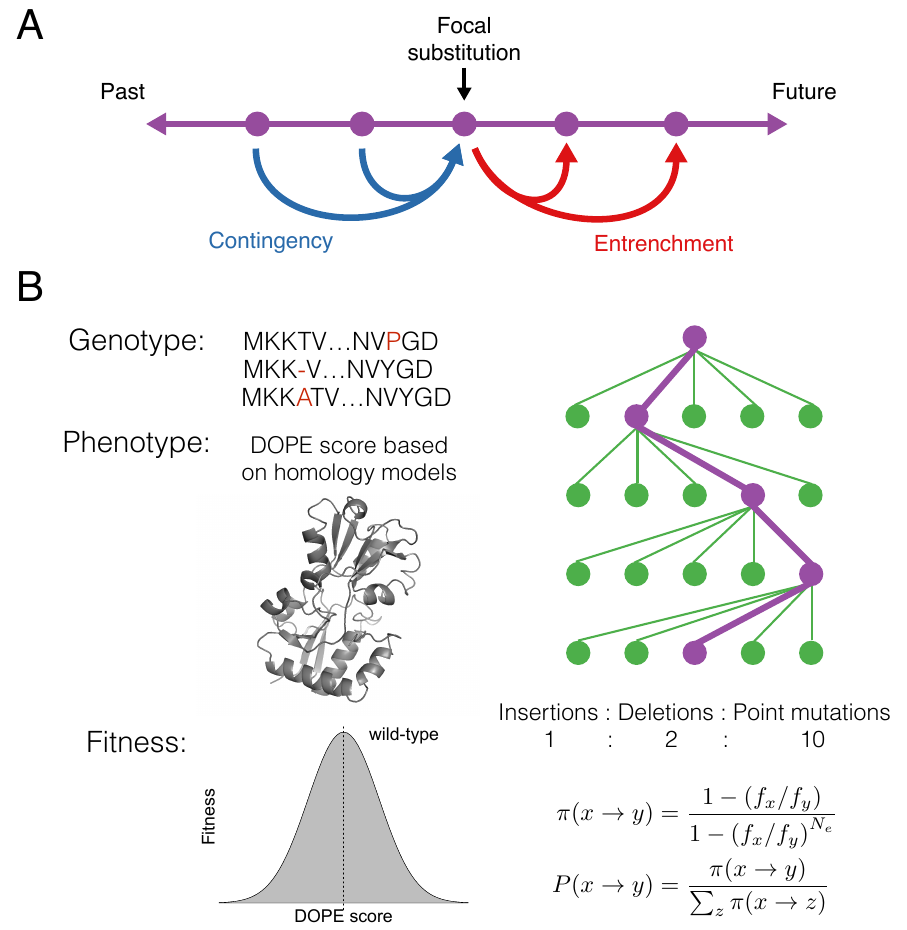}
	\caption{({\bf A}) A schematic model indicating how a focal substitution may be
	contingent on prior 
	substitutions and may constrain future substitutions along an evolutionary
	trajectory, as a result of epistasis.
	({\bf B}) A model of protein evolution under
	weak mutation and purifying selection for native stability. Starting from
	the wild-type sequence of \emph{argT} we propose random one-amino acid 
insertions,
	deletions, and point mutations with relative frequencies $1:2:10$. For each of
	the proposed mutants we compute its predicted stability (DOPE score)
	based on homology models, and its associated fitness relative to the stability
	of the wild-type.
	The population fixes one of the proposed mutants, based on its
	relative fixation probability under the Moran model with population size
	$N_e$. This process is iterated for 50
	consecutive substitutions, which form an evolutionary trajectory. }
	\label{f_concept}
\end{figure}

\begin{figure}[ht]
	        \centering
	\begin{subfigure}
		\centering
		\includegraphics[width=.45\textwidth]{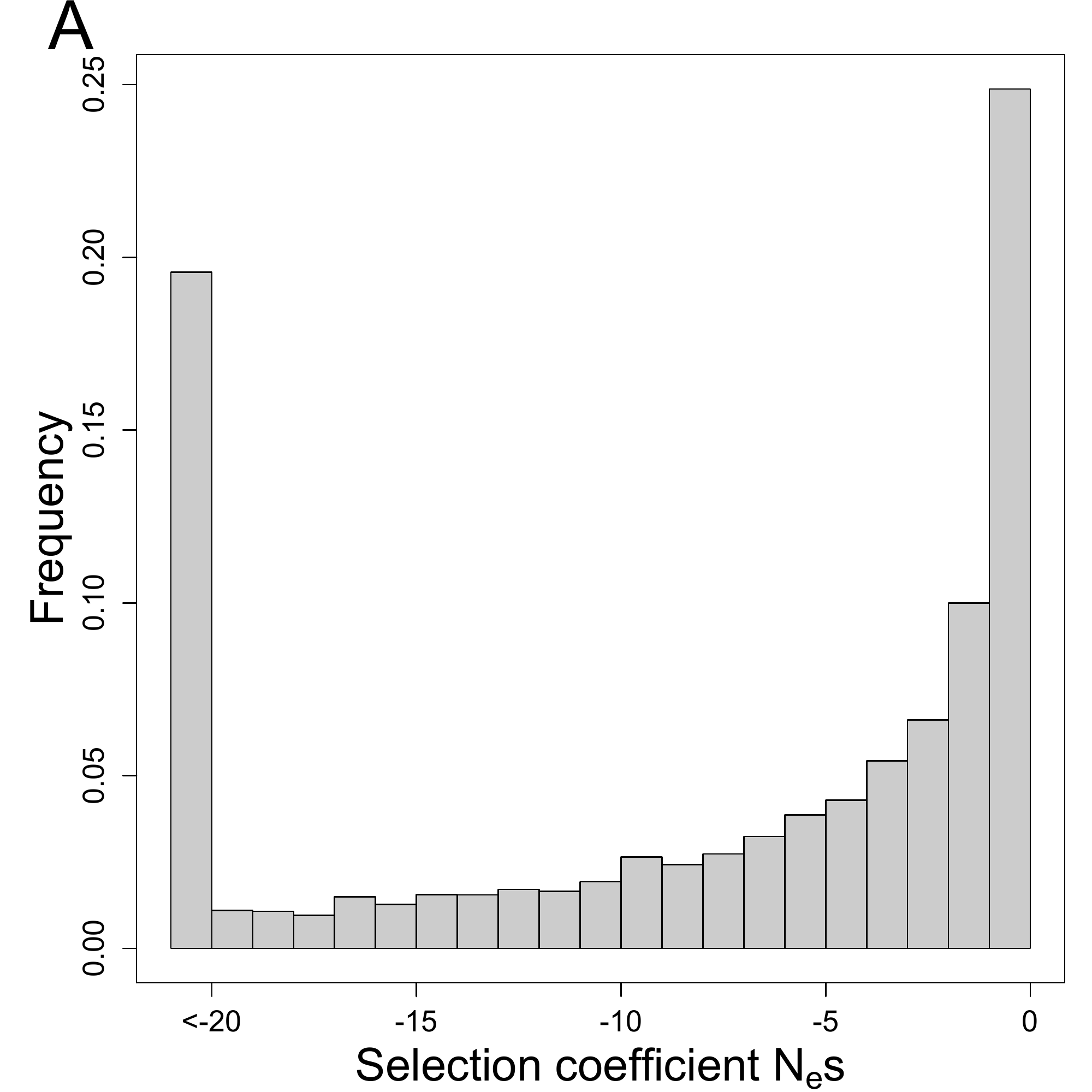}
	\end{subfigure}
	\begin{subfigure}
		\centering
	        \includegraphics[width=.45\textwidth]{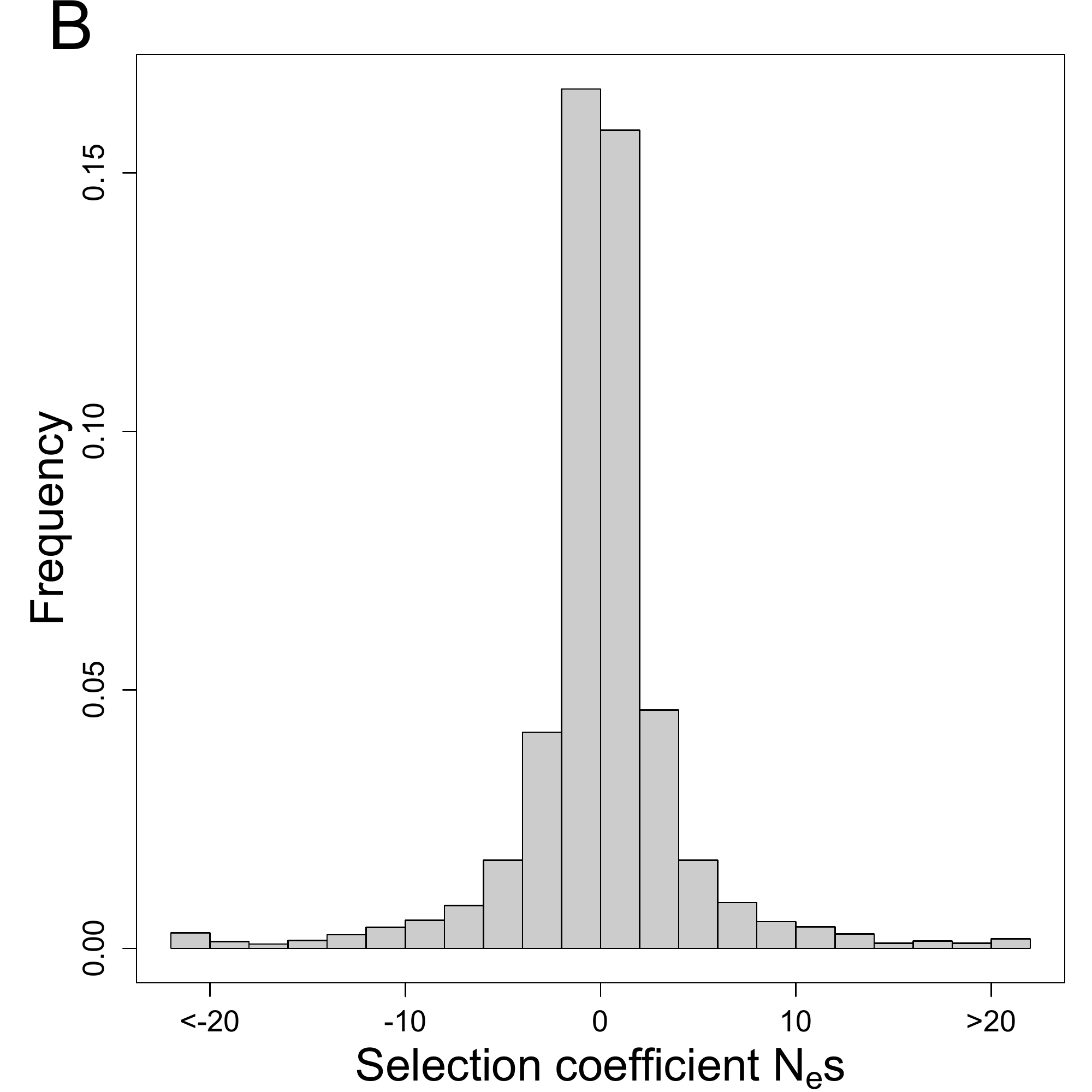}
	\end{subfigure}
\caption{
	({\bf A}) Distribution of selection coefficients
	of all one-step mutations around the wild-type \textit{argT} sequence, based on
	our homology modelling and fitness function. Roughly 25\% of all 
	one-point mutants are nearly-neutral with
	$|N_es|<1$, while roughly 20\% of mutants are practically lethal with $N_es< -20$. 
	({\bf B}) Under
	purifying selection, most substitutions that accrue in our simulations are nearly-neutral. 
	The histogram shows the scaled selection coefficients of all substitutions across
	100 replicate simulated evolutionary trajectories.}
	\label{f_dist_sel_coeff}
\end{figure}

We model purifying selection on protein stability by assuming a Gaussian fitness
function centered around the DOPE score of the wild-type \emph{argT} sequence (Figure \ref{f_concept}B).  
As a result of this form of selection for its native stability, either destabilizing
mutations or over-stabilizing mutations produce variants of \emph{argT} with lower
fitness than the wild-type sequence. (We alternatively consider a fitness function
that penalizes only destabilizing mutations, see below).  We assume an effective population size of
$N_e=10^4$ for the purpose of computing the fixation probabilities of mutants.
The variance of the fitness function is fixed at $10^8$ (arbitrary units) such that roughly $25\%$ of all
possible one-step mutations from the wild-type \emph{argT} sequence have a
scaled selection coefficient $|N_es|<1$ and about 20\% of all mutations are virtually
lethal, $N_es< -20$ (Figure \ref{f_dist_sel_coeff}A).  This choice of fitness function is thus
consistent with experimental data on the distribution of fitness effects of
mutants \cite{Sanjuan:2004dd,EyreWalker:2007dl,Jacquier:2013fx,Bank:2014gx}.

Following Gillespie \cite{Gillespie:1983ws} we implement evolution under weak
mutation as follows.  We initialize the population fixed for the wild-type
\emph{argT} sequence.  At each discrete time-step we propose a set of mutations to
the current sequence, $x$. The proposed mutations include single amino-acid
insertions, single amino-acid deletions, and point mutations with relative
frequencies chosen to roughly match their empirical rates, $1:2:10$ respectively
\cite{Ophir:1997wi,delaChaux:2007fg,Chen:2009dc,TothPetroczy:2013dl}.  We compute
the fixation probability for each of the mutants, $y$, according to the standard
Moran process \cite{Ewens:2004th}: \begin{equation} \pi(x\rightarrow
y)=\frac{1-({f_x}/{f_y})}{1-({f_x}/{f_y})^{Ne}} \end{equation} where $f_x$ denotes
the fitness of genotype $x$, and $\pi(x\rightarrow y)$ denotes the fixation
probability of a mutant genotype $y$ introduced into a population fixed for
genotype $x$.  Next, we let genotype $y$ fix according to its fixation probability
relative to all proposed mutants, \begin{equation} P(x\rightarrow
y)=\frac{\pi(x\rightarrow y)}{\sum_z\pi(x\rightarrow z)}, \end{equation} and we
update the state of the population from sequence $x$ to sequence $y$.  We iterate
this process for a total of 50 discrete time-steps, each corresponding to a
substitution event, so that the final protein sequence is achieved by an
evolutionary trajectory of 50 substitutions starting from the wild-type \emph{argT} 
sequence (Figure \ref{f_concept}B). For simplicity, we use 
the term ``substitution" to refer to the fixation of a
point mutation, insertion, or deletion.
The timescale of our simulations therefore represents roughly 20\% divergence at
the protein sequence level, which is similar to divergences often studied by
comparative sequence analysis. We simulate 100 replicate trajectories and
typically report results on the ensemble average.

\subsection{Homology modeling using \texttt{Modeller}}
Computing a homology model of a mutant involves two steps: alignment and model
building.  For each of the mutants proposed, we generate a structure-dependent
sequence alignment with the wild-type sequence/structure, using the function
\texttt{align2d} in \texttt{Modeller} \cite{Sali:1993ie,Eswar:2008ge}.  This
alignment takes into account the structural properties of the protein in
constraining the sequence alignment.  The homology model for the mutant 
structure
is then built using the \texttt{automodel} function, using High Resolution
Discrete Optimized Protein Energy (DOPE-HR) and normalized DOPE score as
assessment criteria.  The DOPE score of a protein is an atomic distance-
dependent
statistical potential derived from known native structures, and it serves as a
proxy for protein stability \cite{Shen:2006dl}.  The DOPE score takes into account 
the relative
probability of finding two particular atoms at specific distance away from each
other in the structure, given their observed frequencies in native protein
structures in Protein Data Bank \cite{Rose:2011cl}.  
Normalized DOPE is a measure of the
probability that the protein sequence folds to the observed 3D structure, compared
to other possible conformations.  The model is initially optimized with the
variable target function method (VTFM) with conjugate gradient.  Following
VTFM optimization, the model is then refined using molecular dynamics
simulations with simulated annealing \cite{Sali:1993ie}. 

\subsection{Quantifying epistasis, contingency, and entrenchment}
We seek to understand the structure of epistasis between substitutions along
evolutionary trajectories of protein sequences under purifying selection. To
quantify epistasis we use a standard definition for pairs of subsequent mutations,
as well as a natural generalization of this definition for longer trajectories.

Consider first the case in which the population starts at some genotype $S_0$ with
fitness $f_0$.  Upon fixation of the first substitution the population moves to
genotype $S_{0,1}$ with fitness $f_{0,1}$. Upon fixation of the second substitution the
population moves to genotype $S_{0,1,2}$ with fitness $f_{0,1,2}$. In the absence of
the first mutation, the second mutation would have moved the population to
genotype $S_{0,2}$ with fitness $f_{0,2}$.  The standard measure of epistasis between
these two substitutions is defined as 
\begin{align} \label{Estandard}
\begin{split}
E&= \Big[ \log (f_{0,1,2})-\log(f_{0})\Big]\\ 
& \quad \quad - \left(\Big[\log(f_{0,1})-\log(f_{0}) \Big]+\Big[\log(f_{0,2})-\log(f_{0}) \Big]\right).
\end{split}
\end{align}
Writing the definition in this way suggests that we
view epistasis as the deviation between the fitness effect of the double mutant
and the sum of the fitness effects of the single mutants.

This definition of epistasis can alternatively be interpreted
in terms of the order in which 
substitutions occurred along the evolutionary trajectory. For instance, in the above scenario 
mutation $1$ fixes before mutation $2$ and it therefore has fitness effect $\log(f_{0,1})-
\log(f_0)$. However, we can also ask: what {\em would} the fitness effect of 
mutation $1$ have been had the two mutations fixed in the opposite order? In this alternative 
scenario, the fitness effect of mutation $1$ would be $\log(f_{0,1,2})-\log(f_{0,2})$. 
The standard definition of epistasis between a pair of mutants can be 
re-written as the difference between these two fitness effects:
\begin{align}
\label{E}
E&=\Big[\log(f_{0,1,2})-\log(f_{0,2})\Big]-\Big[\log(f_{0,1})-\log(f_0)\Big].
\end{align}
Thus, the standard measure of epistasis can be seen as a measure of how much larger the 
fitness effect of the first substitution would be if the order of the 
two substitutions were reversed.

This interpretation of epistasis 
in terms of substitution order suggests a natural generalization, which will allow
us to quantify epistasis in longer evolutionary trajectories. Consider a trajectory starting at the wild-type 
sequence and then subsequently fixing mutations $1, 2, 3, \ldots, n$. For any 
mutation $i$, we can ask how much larger the fitness effect of mutation $i$ would 
have been under the alternative trajectory in which mutation $i$ is removed 
from position $i$---where it actually occurred---and instead inserted at some other 
position $j$ along the trajectory. More formally, in such a trajectory
we define the following measure to quantify epistasis between substitutions $i$
and $j$:
\begin{align}
E_{(i,j)} =
\begin{cases}
\Big[\log(f_{0,1,\ldots,j-1,i})-\log(f_{0,1,\ldots,j-1})\Big]-\\
\hfill \Big[\log(f_{0,1,\ldots,i})-\log(f_{0,1,\ldots,i-1})\Big] ,
 & \text{ for } i\geq j \vspace{1ex} \\
\Big[\log(f_{0,1,\ldots,j})-f_{0,1,\ldots,i-1,i+1,\ldots j})\Big]-\\
\hfill \Big[\log(f_{0,1,\ldots,i})-\log(f_{0,1,\ldots,i-1})\Big],
& \text{ for } i <j,
\end{cases}
\end{align}
It is easy to verify that $E_{(i,i+1)}$ reduces to the standard 
measure of epistasis between two subsequent substitutions.

This generalized definition of epistasis allows us to define precisely what we
mean by contingency and entrenchment. A substitution is contingent on previous
substitutions if it is more likely to fix as a result of the substitutions that
preceded it. More precisely, for $i>j$ we define substitution $i$ to be contingent
on the preceding substitutions $j,\ldots,i-1$ if $E_{(i,j)}<0$. The condition $E_{(i,j)}<0$
means that substitution $i$ is relatively more beneficial when it actually
occurs than it would have been had it occurred at some earlier time-step, $j$. 
Likewise, we say that a substitution $i$ is
entrenched by subsequent substitutions if it becomes relatively more deleterious to revert
as a result of the subsequent substitutions.  More precisely, for $i<j$ we say
a substitution $i$ is entrenched by subsequent substitutions $i+1,
\ldots, j$ if $E_{(i,j)}>0$. The condition $E_{(i,j)}>0$ means that the effect of
reverting substitution $i$ at time $j$ is relatively more deleterious than it
would be to revert substitution $i$ immediately after it initially occurred.

\section{Results}

\subsection{Mutational effects on protein stability}
Random mutations in a protein-coding sequence are known to typically
destabilize the protein structure \cite{Arnold:2001uu,Taverna:2002to,Depristo:2005jc,Bloom:2006ch,Tokuriki:2007ij,Bloom:2007jj,Zeldovich:2007hv,Jacquier:2013fx}.
Thus, if protein evolution proceeded solely via random substitutions, without any
selection, we would expect a decrease in protein stability over time.
However, under purifying selection to maintain a given degree of thermodynamic
stability, strongly
destabilizing or over-stabilizing mutations 
will both have low fitness and correspondingly low fixation probability, so that 
the only mutations that substitute will tend to produce stabilities similar to that of 
the
wild-type sequence.

We simulate the evolution of the \emph{argT} protein sequence under selection for 
its native stability, computing homology models, stability scores,
and fixation probabilities of mutants as described above. Proposed mutations 
included point mutations, insertions, and deletions. 
As expected, the stability of the protein remained fairly constant along these evolutionary trajectories with almost equal
number of stabilizing and destabilizing mutations substituted over time (Binomial 
test, $p=0.7$). Moreover, these substitutions were typically nearly-neutral
(Figure \ref{f_dist_sel_coeff}B), such that the fitness of the protein 
decreased by only $\sim0.05\%$ on average after 50 substitutions.
We found no significant difference in the distribution
of fitness effects of stabilizing versus destabilizing mutations that fixed along 
these evolutionary trajectories under purifying selection (t-test, $p=0.17$).

By contrast, when we simulate protein sequence evolution via the fixation of 
random indels or point mutations -- that is, without any selection at all
-- then the stability for the native structure decreases along evolutionary 
trajectories, as illustrated by the ensemble mean trajectory shown in
Figure \ref{f_dope}.  And in the absence of selection, substitutions are more often 
destabilizing than stabilizing (Binomial test, $p<10^{-15}$).  These
results suggest that the homology models and DOPE metric are consistent with 
empirical studies on the effects of random mutations on protein stability
\cite{Arnold:2001uu,Taverna:2002to,Zeldovich:2007hv,Ashenberg:2013gd,Jacquier:2013fx}.

\subsection{Epistasis along evolutionary trajectories:~contingency} We quantified
the structure of epistasis between substitutions along evolutionary
trajectories of \emph{argT} sequences simulated under purifying selection.
We used a
generalized definition of epistasis $E_{(i,j)}$ that applies to any pair of
substitutions $i$ and $j$ along a trajectory (see \emph{Methods}). We first
studied the degree of contingency between substitutions in these trajectories.
For $i>j$ we say that substitution $i$ is contingent on 
the preceding substitutions $j, \ldots,
i-1$ if the condition $E_{(i,j)}<0$ holds. This contingency condition 
means that substitution $i$ is
relatively more beneficial at the time of its actual occurrence than it would have
been had it occurred at some earlier step, $j$. 

We find that substitutions in \emph{argT} under purifying selection are highly epistatic 
and they tend to be contingent on 
earlier substitutions.
Figure \ref{f_sel_cont_ent}A (left side) illustrates this phenomenon by focusing
on contingency between the substitutions that occur at step $i=25$ and
the substitutions that occur at earlier steps $j<25$, among an ensemble of replicate
evolutionary trajectories.
The mean epistasis measure $E_{(25,j)}$ is 
significantly less
than zero for each step $j<25$ (t-test, $p<10^{-4}$ for all $j$) -- indicating that
the substitutions that fix at step $i=25$ are highly contingent on earlier substitutions.

There is a subtlety associated with the contingency condition $E_{(i,j)}<0$, which
compares the selection coefficient of substitution $i$ when it fixes versus the
selection coefficient of the same mutation had it fixed at some earlier step, $j$.  These two selection
coefficients can each be negative or positive. When $E_{(i,j)}<0$, it
means substitution $i$ is ``relatively more beneficial" at time $i$ compared to
at a prior time; this includes the possibility that substitution $i$
is in fact deleterious, but less deleterious at the time of its actual fixation compared to having
fixed at some earlier time. In practice, in simulations under purifying
selection most of the mutations that fix along the evolutionary trajectory
are neutral or nearly-neutral at the time of their fixation (Figure \ref{f_dist_sel_coeff}B). And so in these 
simulations the condition $E_{(i,j)}<0$
typically means that 
substitution $i$ would have been
strongly deleterious had it occurred at earlier step $j$. 

This form of contingency is illustrated in Figure \ref{f_sel_cont_ent}B, which
compares the selection coefficients of the mutations that fix at step $i=25$ with the
selection coefficients of the same mutations had they been introduced at earlier
steps $j<i$ along their evolutionary trajectories. As the figure shows,
substitutions at step $i=25$ are typically neutral or nearly-neutral; but the same
mutations would often be strongly deleterious were they introduced in prior
genetic backgrounds. Hence, the substitutions at step $i=25$ are contingent on
earlier substitutions: they typically would not have fixed without the presence of
those earlier substitutions.

More generally, when considering all pairs of substitutions $j<i$ in our
simulations under purifying selection we find that $\sim79\%$ of them exhibit
$E_{(i,j)}<0$, with a mean $N_eE=-14.1$.
In other words, the great majority of substitutions that fix are contingent on
earlier substitutions.
Moreover, this pattern holds
separately for each of the three types of focal mutations: point mutations (78\% with $E_{(i,j)}<0$),
insertions (84\% with $E_{(i,j)}<0$), and deletions (86\% with $E_{(i,j)}<0$). These results imply that,
even under purifying selection for stability, the mutations that fix during
the evolution of a protein sequence are highly contingent on the history of prior substitutions. 
As a result of
this contingency,
the long-term evolutionary trajectory cannot be predicted by assaying the
phenotypic effects of these mutations introduced into the wild-type sequence alone.

\subsection{Epistasis along evolutionary trajectories:~entrenchment}
We have shown that mutations that fix under purifying selection are often contingent upon 
earlier substitutions.
Now we ask the converse question: what is
the effect of later substitutions on the fitness effects of substitutions that have 
already fixed?
In particular, we ask whether mutations which are nearly-neutral when they fix
subsequently become deleterious to revert later in the trajectory -- a question which 
is similar to the one studied by Pollock et al \cite{Pollock:2012ge}.

A positive value of $E_{(i,j)}$ for $j>i$ means that reverting substitution $i$ in a 
later background
$i+1,\ldots,j$ is relatively more deleterious than reverting it immediately after it 
fixes in the population, indicating
\emph{entrenchment} of substitution $i$.

\begin{figure}[ht]
	        \centering
	\begin{subfigure}
		\centering
		\includegraphics[width=.45\textwidth]{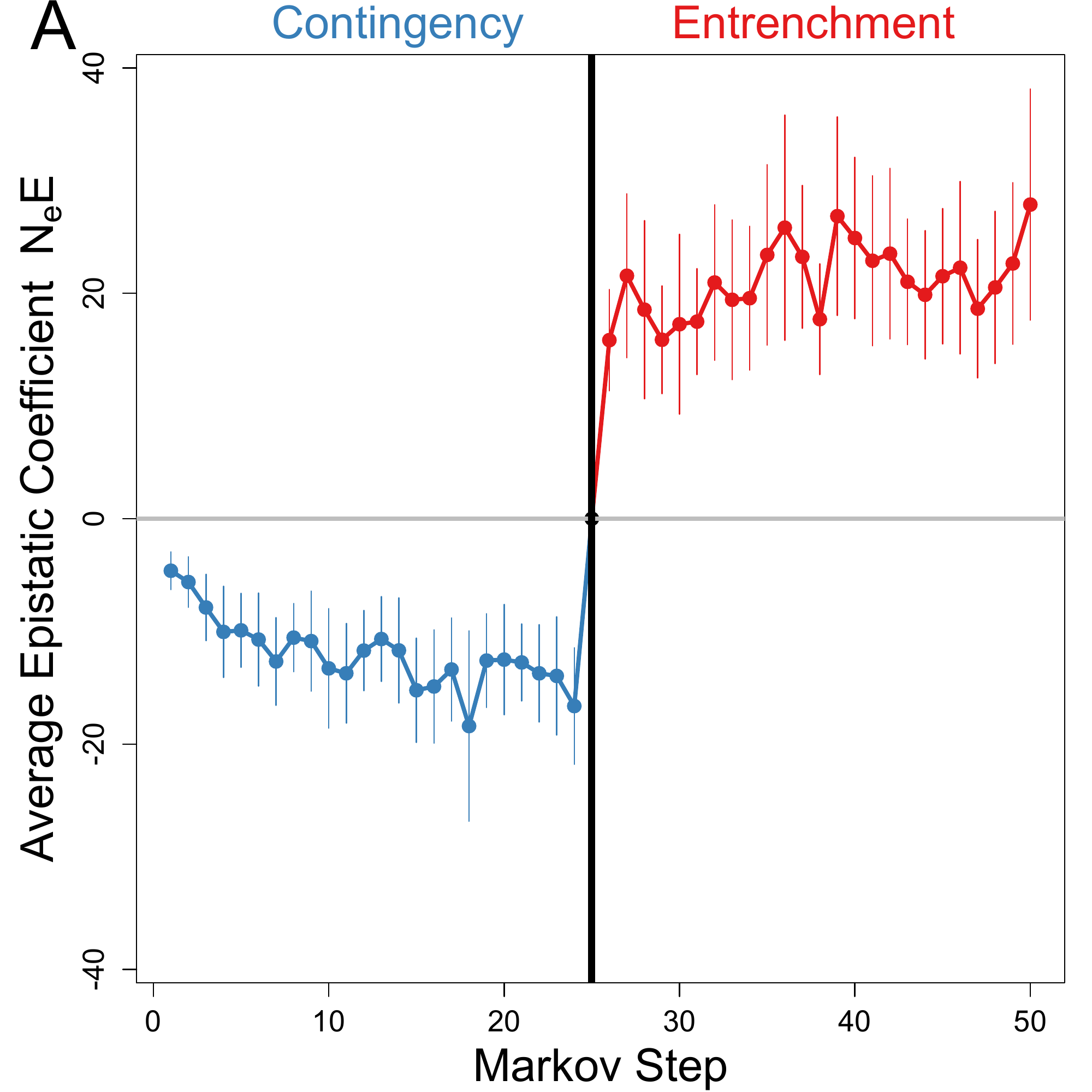}
	\end{subfigure}
	\begin{subfigure}
		\centering
	        \includegraphics[width=.45\textwidth]{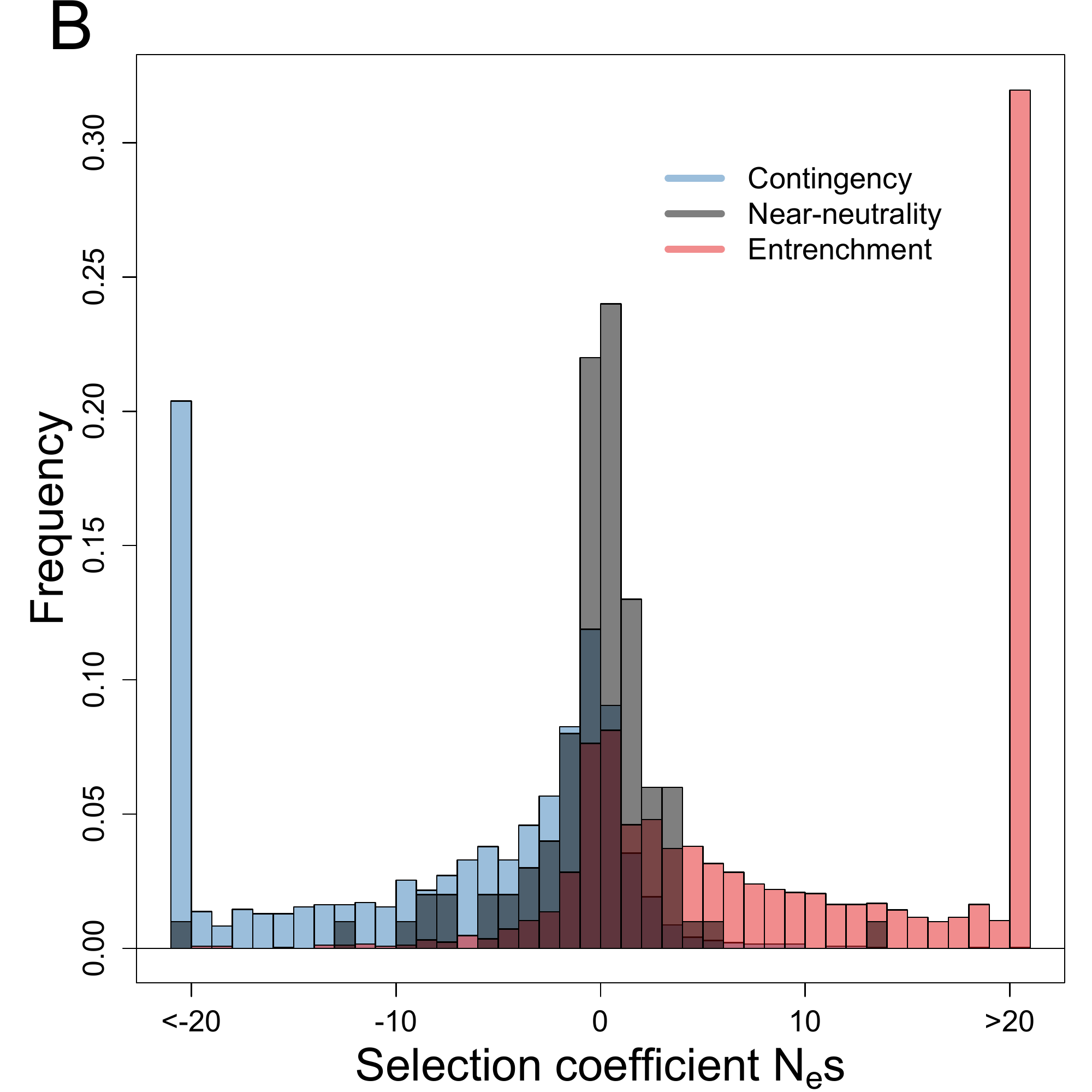}
	\end{subfigure}
\caption{
({\bf A})Substitutions that accrue under purifying selection are highly
epistatic: they exhibit both contingency with earlier substitutions
and entrenchment due to later substitutions.
The figure indicates
the fitness effect of substitutions that fixed at step $i=25$
in earlier (contingency $j<25$)
or later (entrenchment $j>25$) backgrounds.
Under purifying selection, the epistatic coefficients $E_{(25,j)}$ are
significantly
less than zero for all $j<25$ and significantly greater than zero for
all $j>25$.
Thus, substitutions that are nearly-neutral
when then fix are highly
contingent on earlier substitutions; and they become deleterious to revert as
later substitutions accrue.
Vertical bars indicate $\pm2$ SE around the ensemble mean of
   100 replicate simulated populations. ({\bf B}) Distribution of
scaled selection coefficients ($N_e s$) for substitutions that fix at
step $i=25$. The gray histogram shows the distribution of selection
coefficients of these mutations at the time that they fix
(``near-neutrality"); the blue histogram shows the distribution of
selection coefficients for the same mutations if they were introduced
in backgrounds $j=0, \ldots, 25$
(``contingency"); and the red histogram shows the distribution of
selection coefficients for the same mutations if they were introduced in
backgrounds $j=26, \ldots, 49$,
assuming that substitution 25 had never occurred (``entrenchment").
}
	\label{f_sel_cont_ent}	
\end{figure}

We find that substitutions under purifying selection are typically 
entrenched by later substitutions.  Figure \ref{f_sel_cont_ent}A (right side) illustrates this phenomenon
by focusing
on entrenchment of substitutions that occur at time $i=25$ by
substitutions that occur at later time-points $j>i$ along the same evolutionary trajectories.
The
mean entrenchment coefficient $E_{(25,j)}$ is significantly greater than zero
for each subsequent step $j>25$ (t-test, $p<10^{-4}$ for each
$j$). Furthermore, 
as Figure \ref{f_sel_cont_ent}B shows, most of the substitutions that fix at step $i=25$ are nearly-neutral,
whereas reverting the same mutations from later genetic backgrounds is typically highly 
deleterious (Figure \ref{f_sel_cont_ent}B). 

More generally, when considering all pairs of substitutions $i<j$ under purifying
selection, epistatic values $E_{(i,j)}$ for $j>i$ are significantly greater than zero on
average (t-test, $p<10^{-15}$) with a mean $N_{e}E_{(i,j)}=17.95$,
indicating a tendency for later substitutions to entrench earlier
substitutions.
In particular, $\sim81$\% 
pairs of such substitutions exhibit positive values $E_{(i,j)}$, a pattern that also
holds within each of the three classes of focal substitutions $i$: point mutations (81\%
with $E_{(i,j)}>0$),
insertions (79\% with $E_{(i,j)}>0$), and deletions (75\% with $E_{(i,j)}>0$).

\begin{figure}[ht]
        \includegraphics[width=.45\textwidth]{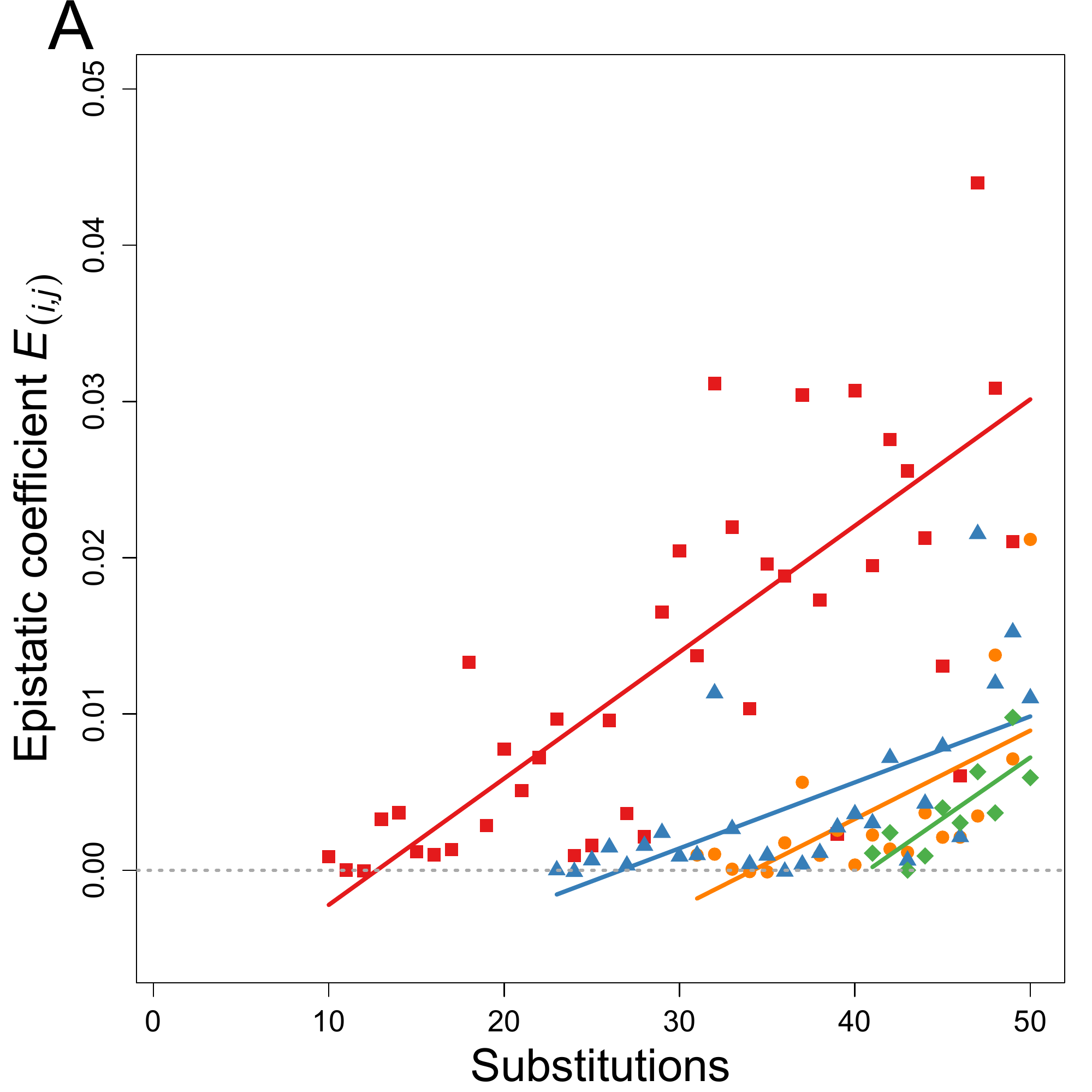}
\caption{Substitutions that fix under purifying selection are 
increasingly entrenched by subsequent
substitutions. The figure indicates
the fitness effect of reverting four illustrative substitutions (occurring at
$i=10, 23, 31, 41$ in independent trajectories), each represented by a
unique color, in independent evolutionary trajectories at subsequent steps
along their respective trajectory.
The degree of entrenchment $E_{(i,j)}$ of each of these four focal substitutions $i$ 
by subsequent
substitutions $j>i$ increases along the trajectory, so that the slope of
$E_{(i,j)}$ is positive as $j$ increases. 
In other words, reverting a substitution becomes increasingly deleterious as
subsequent substitutions accrue.
Solid lines indicate the best-fit linear regression of the epistatic coefficient $E_{(i,j)}
$ on substitution number $j$ for $j>i$.
Aside from these four illustrative examples, the slopes of $E_{(i,j)}$ are positive, 
on average, across all
substitutions in all our simulated evolutionary trajectories (one-tailed
t-test, $p<10^{-15}$). }
	\label{f_ent_traj}	
\end{figure}

Moreover, the degree to which a substitution becomes entrenched by epistasis
tends to increase with each subsequent substitution. Figure \ref{f_ent_traj} illustrates this
phenomenon, by showing the amount of epistasis between a focal substitution $i$ 
and
each subsequent substitution $j>i$ along an individual evolutionary trajectory.
The figure shows four illustrative cases where we observe that the focal
substitution $i$ becomes increasingly deleterious to revert as subsequent
substitutions accumulate, as demonstrated by the positive slope of $E_{(i,j)}$ as $j$ increases. 
Indeed, this slope is significantly positive, on average, across all
substitutions $i$ in our simulations (one-tailed t-test, $p<10^{-15}$); and 
60\% of substitutions exhibit positive slopes, indicating a modest (but statistically 
highly significant) tendency for entrenchment to increase over time
(see also Figure \ref{f_epist_mat}).  
Thus, even under purifying selection, we find that  protein-coding substitutions are 
rendered ``irreversible" by
subsequent substitutions and that the degree of irreversibility often increases with time.

\subsection{Epistasis between consecutive substitutions} 
We have shown that the selection coefficient of a given substitution is contingent 
on prior
substitutions and becomes entrenched by subsequent substitutions, constraining 
evolution against reversions as
time proceeds.  However, does epistasis constrain the
paths available to evolution on shorter time scales as well -- that is, between 
immediately consecutive substitutions?

To address this question we consider an evolutionary trajectory starting at 
genotype $A$
followed by subsequent substitutions $B$ and $C$, producing the trajectory of 
genotype
$A\rightarrow AB\rightarrow ABC$. We ask how likely is the observed path compared 
to the alternative
path $A\rightarrow AC\rightarrow ACB$?
We calculate the probability of seeing one path versus the other based on their fixation 
probabilities
\begin{equation}
\begin{split}
&P(A,BC)=\\ & \quad \frac{\pi(A\rightarrow AB)\pi(AB\rightarrow ABC)}{\pi(A\rightarrow AB)\pi(AB\rightarrow ABC)+\pi(A\rightarrow AC)
\pi(AC\rightarrow ABC)}.
\end{split}
\end{equation}
A value $P(A,BC)>1/2$ indicates that the actual path taken during evolution ($A
\rightarrow AB\rightarrow ABC$) is more favorable than 
the alternative path ($A\rightarrow AC\rightarrow ACB$), and vice versa.

\begin{figure}[ht]
		\includegraphics[width=.45\textwidth]{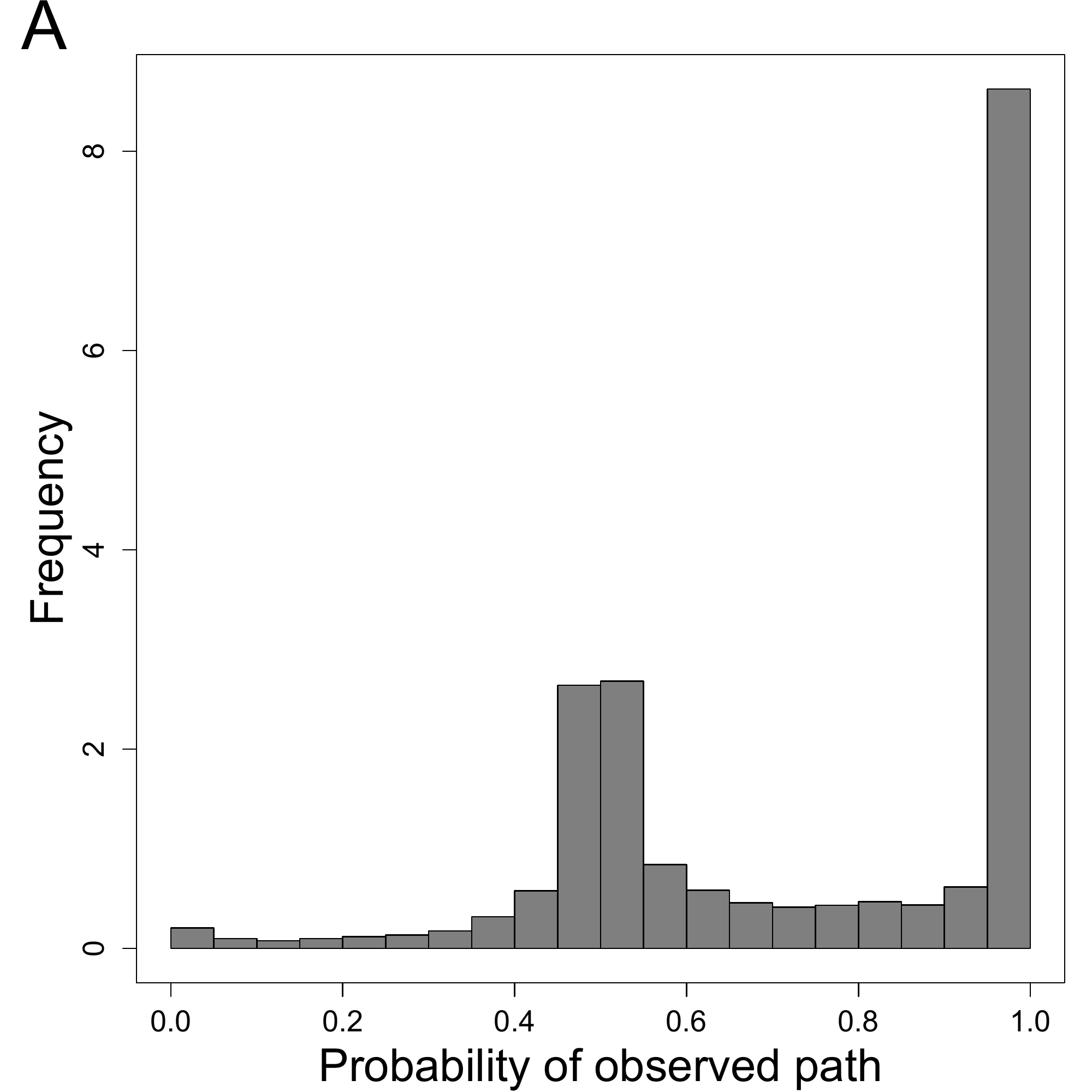}
	\caption{Purifying selection constrains paths available to evolution. The figure 
shows the probability of fixing two consecutive
	substitutions ($B$ and $C$) in their observed order ($A\rightarrow AB
\rightarrow ABC$) compared to the reversed order ($A\rightarrow AC\rightarrow 
ACB
$).
	({\bf A}) Under purifying selection, for about 40\% of pairs of substitutions, 
	the observed path is more than 50 times as likely as the alternate path, indicating
	that many substitutions are highly conditional on the immediately preceding one. 
	}
	\label{f_relprob_hist}	
\end{figure}

We calculated the relative probabilities of alternate paths for all pairs of
consecutive substitutions in the ensemble of evolutionary trajectories. These
probabilities show an interesting bimodal pattern (Figure \ref{f_relprob_hist}).
While for some substitutions the actual and alternative paths were approximately
equally probable, for $\sim40$\% of consecutive pairs the actual path was more
than 50-times as likely as the alternate path.  This indicates a high degree of
epistasis between a large portion of consecutive substitutions, and it shows that
many substitutions are conditional on the presence of the immediately preceding
substitution. Thus, even over short time-scales, epistasis plays a large role in
shaping the mutation paths taken by evolution under purifying selection.

\subsection{Frequency of positive, negative, and sign epistasis}
Adaptive substitutions arising across a bacterial genome tend to be negatively 
epistatic
with each other -- that is, their combined effects are less beneficial than
expected under independence
\cite{Kryazhimskiy:2009er,Chou:2011dk,Khan:2011ik,Kryazhimskiy:2011hu, Wiser:2013ch}.
By contrast, empirical studies on individual proteins have typically found that
adaptive substitutions are sign epistatic -- that is, substitutions increase
protein enzymatic activity in one background but decrease it in other backgrounds
\cite{Weinreich:2006ig,Bridgham:2009jv,Natarajan:2013gi}. Recent attempts 
to reconcile these patterns of
epistasis \cite{Draghi:2013kg,Greene:2014db} have shown that
the predominant form of epistasis changes systematically along evolutionary
trajectories, shifting from negative epistasis, early in adaptation, to sign
epistasis, as populations reach local fitness maxima.

To compare our results on protein evolution with this literature, we have quantified 
the
frequencies of positive, negative, sign, and reciprocal sign epistasis at each
step in our
simulations. 
We say that a pair of substitutions, $1$ and $2$, shows reciprocal sign epistasis if both $\text{sign}\left[\log\left(\frac{f_{0,1,2}}{f_{0,2}}\right)\right] \ne 
\text{sign}\left[\log\left(\frac{f_{0,1}}{f_{0}}\right)\right]$ and $\text{sign}
\left[\log\left(\frac{f_{0,1,2}}{f_{0,1}}\right)\right] \ne \text{sign}\left[\log\left(\frac{f_{0,2}}{f_{0}}\right)\right]$; 
we say that the pair shows sign epistasis if only one of these inequalities hold.
For pairs of substitutions that exhibit neither sign nor reciprocal sign
epistasis, we classify epistasis as positive if $E>0$ and negative if $E<0$.

The relative frequencies of these four forms of epistasis do not change
substantially along evolutionary trajectories under purifying selection (Figure
\ref{f_epist_type}A).  Throughout time we find that sign, reciprocal-sign and
positive epistasis dominate, whereas negative epistasis is only observed a small
fraction of the time (Figure \ref{f_epist_type}A).  This pattern can be understood
intuitively in terms of our observation above that many substitutions are
contingent upon the immediately preceding substitution. Geometrically, this form
of contingency corresponds to the movement of a population along a fitness ridge
($f_0 \rightarrow f_{0,1}\rightarrow f_{0,1,2}$, where $f_{0,2}$ has low fitness
compared to $f_0$, $f_{0,1}$ and $f_{0,1,2}$). 
Such a ridge-traversal can produce sign, reciprocal-sign, and positive epistasis; but
it will never produce negative epistasis.

To complement our studies of purifying selection we also studied epistasis in
evolutionary trajectories of proteins simulated under directional selection.  To
do so we initiated populations at the wild-type sequence \emph{argT} sequence, but
shifted the optimum value of the DOPE score away from that of the wild-type
sequence.  The optimum DOPE score was chosen so that it lies 100 $N_es$ units away
from the wild-type DOPE score.  We again simulated an ensemble of populations
evolving under weak mutation, as above, and we calculated the degree of epistasis
at each time-step.  However, instead of running each evolutionary trajectory for a
fixed number of steps, we only let adaptive mutations fix and stopped each
simulation when it reached a local optimum, as in \cite{Draghi:2013kg}. 
The pattern of epistasis under adaptive evolution is very different from the
pattern we observe under nearly-neutral evolution (Figure \ref{f_epist_type}B).
Unlike under nearly-neutral evolution, the nature of epistasis in  these adaptive
trajectories changes substantially through time. The adaptive trajectories are
initially dominated by negative epistasis, which was rare under nearly-neutral
evolution. However, sign epistasis begins to dominate as populations move towards
a phenotypic optimum. These observations confirm, in a mechanistically-derived
protein stability landscape, the trends previously reported for mathematical
fitness landscapes and for RNA-folding landscapes \cite{Draghi:2013kg}. 

\subsection{Evolution of protein length}

Whereas comparative sequence analyses are typically limited to studying selection
pressures on point mutations \cite{Liberles:2012ir,Breen:2012fd}, 
our framework allows us to study the consequences of
insertions and deletions as well. As a result, we can ask questions about the
determinants of protein length variation over evolutionary timescales.

\begin{figure}[ht]
	        \centering
	\begin{subfigure}
		\centering
		\includegraphics[width=.45\textwidth]{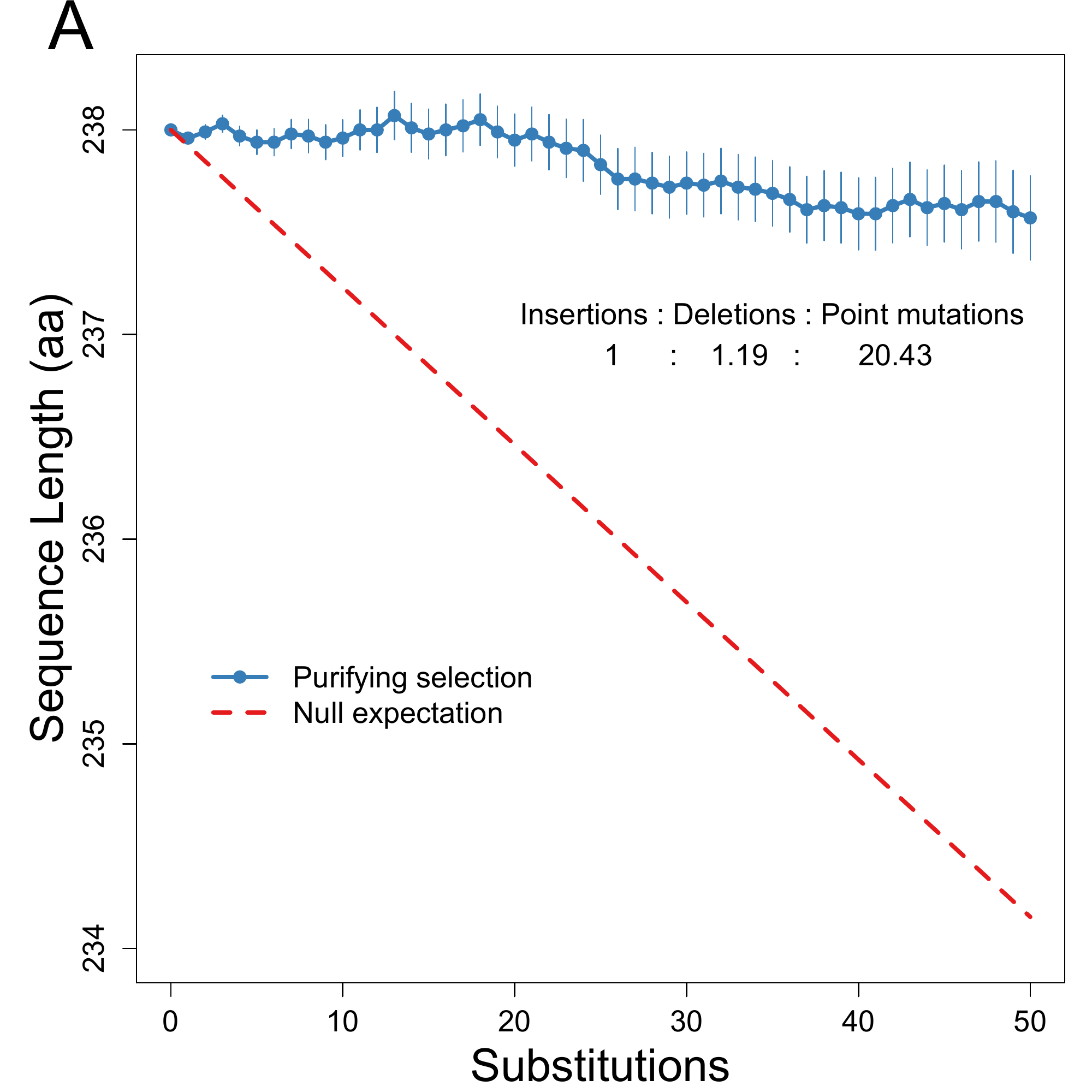}
	\end{subfigure}
	\begin{subfigure}
		\centering
	        \includegraphics[width=.45\textwidth]{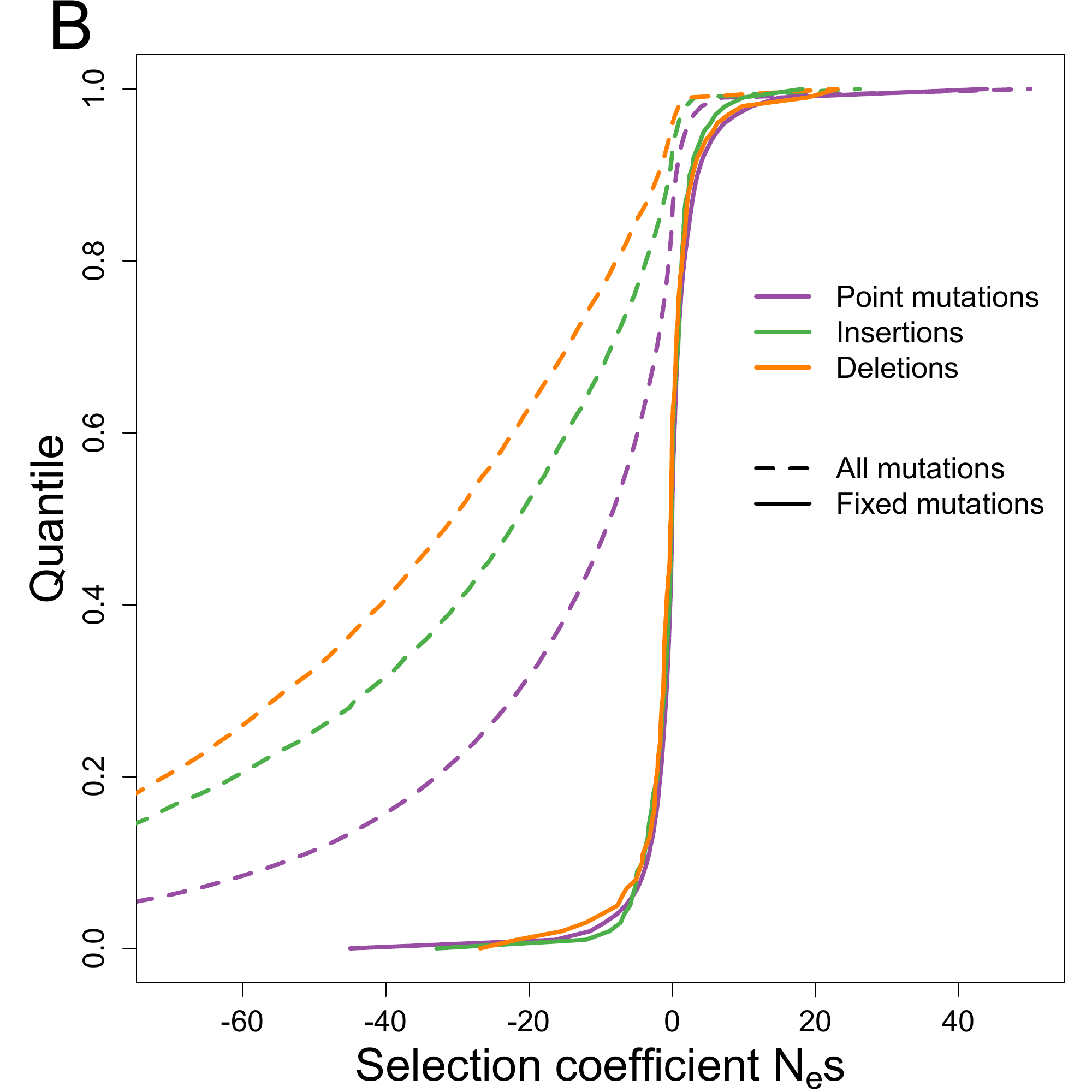}
	\end{subfigure}
	\caption{Protein sequence length remains roughly constant under purifying
	selection. ({\bf A}) In the absence of any selection, the sequence length is 
expected
	to decrease by $\sim3.8$ aa after 50 substitution events
	based on rates of proposed insertions, deletions and point mutations at each
	step. By contrast, under purifying selection  the sequence length decreases 
by
	only $\sim0.5$ aa, on average, after 50 substitutions. Protein length
	is preserved because purifying selection elevates the substitution rate of 
point
	mutations relative to indels; and also decreases the substitution rate of 
deletions
	relative to insertions. Vertical bars indicate $\pm1$ SE around the ensemble 
mean of
	100 replicate simulated populations. ({\bf B}) The distribution of selection 
coefficients
	for random (dotted lines) and fixed (solid lines) insertions, deletions and
	point mutations under purifying selection. Random 
	point mutations are generally less deleterious than random insertions or 
deletions.}
	\label{f_glen_sel}	
\end{figure}

At each step of our evolutionary simulation, we propose insertions, deletions, and
point mutations with relative frequencies of $1:2:10$, and we let one of these
mutations substitute according to its predicted effect on protein stability and
corresponding relative fixation probability in a finite population (see
\emph{Methods}).  In the absence of selection, the length of the simulated protein
is expected to decrease on average by $\sim3.8$ amino acids following 50 such
substitution events, based on proposed rates of insertions, deletions, and point
mutations (Figure \ref{f_glen_sel}A).  In the presence of purifying selection, by
contrast, the length of protein sequence remains roughly constant, decreasing on
average by $<0.5$ aa after 50 steps (Figure \ref{f_glen_sel}A).  The long-term
preservation of protein length is caused by two facts: first, the frequency of
point mutations that fix relative to indels is roughly three times the frequency
at which they are proposed ($\sim10:1$ instead of $10:3$, Figure
\ref{f_glen_sel}A), which retards the rate of change to protein length; second,
the frequency of insertions that fix is only slightly less than that of deletions
that fix ($1:1.2$ ensemble mean), even though they are proposed half as often as
deletions. As a result, there is a very slight tendency to decrease protein
length, by roughly one amino for every 100 substitution events.

The fixation rates of insertions, deletions, and point mutations, relative to the
rates at which these mutations are proposed, reflect differences in the typical
effects of such mutations on protein stability and corresponding differences in
their selection coefficients.  In particular, random point mutations have the
smallest absolute effect on protein stability and fitness, on average, followed by
random insertions, and finally random deletions (t-test, $p<10^{-15}$ for all pairs)
(Figure \ref{f_glen_sel}B).  By contrast, there is no significant difference in the mean 
scaled selection coefficients, $N_eS$, of the mutations that actually substitute in these
simulations as a function of mutation type (t-test, $p>0.5$ for all pairs). 

These results suggest that when proteins evolve under purifying selection for
stability their lengths should remain fairly constant over time scales up to
$\sim20\%$ sequence divergence, despite differences in the
spontaneous rates of random insertions and deletions.  These results are
consistent with broad empirical trends in homologous protein length variation
across divergent species \cite{Wang:2005bq,Xu:2006cl}.
Moreover, these results imply that genes under stronger purifying selection should
exhibit a higher ratio of fixed point mutations to indels, compared to genes under
weak purifying selection; and that this ratio should again be higher in organisms 
with
larger effective population sizes -- predictions that are both consistent with 
empirical observations
\cite{TothPetroczy:2013dl,Chen:2009dc}.

\section{Robustness of simulation results}

All the results reported above are based on 100 replicate simulations of
\emph{argT} evolution under weak mutation. Each simulation was initiated at the 
wild-type
sequence, and then proceeded for 50 discrete steps corresponding to 50 
substitution events.
Instead of proposing all possible mutations at each
step, however, for reasons of computational tractability we
proposed only one insertion, two deletions, and ten point mutations.  

In order to verify that our results are not influenced by the relatively small
sample of mutations proposed at each step we ran a smaller set of shorter
simulations (20 replicates, each for 20 substitutions), proposing in this case 20
different insertions, 40 deletions, and 200 point-mutations at each step.
Our results remained qualitatively unchanged.  In particular, as Figure \ref{f_epist_low_high} 
and Figure \ref{f_sel_low_high} show, 
even with 20-fold higher rate of sampled mutations, we found
the same mean selection coefficients for proposed insertions, deletions, and
point mutations (t-test, $p>.6$ in each category); we still find that over 75\% of
substitutions $i$ are entrenched by later substitutions $j$; and
we still find that 60\% of substitutions are \emph{increasingly} entrenched by 
subsequent
substitutions. 
In addition, the relative frequencies of the various forms of epistasis also remain 
the same
under higher number of proposed mutations at each step ($\chi^2\text{-test, } 
p>0.7$).
Thus, all of our qualitative and even quantitative results are
robust to the number of mutations sampled at each step in the weak-mutation
simulations.

Our model of purifying selection assumes that over-stabilizing mutations are as
deleterious as destabilizing mutations, so that only the wild-type stability has
optimal fitness.  However, several studies have shown that over-stabilizing
mutations are often neutral under stabilizing selection
\cite{Arnold:2001uu,Bloom:2006ch,Zeldovich:2007hv,Chen:2009jb}.  In order
to understand how our results depend on assumptions about stabilizing mutations 
we considered an alternative fitness landscape in which \emph{argT} sequences more
stable than the wild-type are just a fit as the wild-type (Figure \ref{f_alt_model}).  We ran a
smaller set of simulations (20 replicate trajectories, each for 50 substitutions)
on this alternative fitness landscape, and we found that our results remain
qualitatively unchanged.  In particular, the stability of the evolved protein did
not change significantly from that of wild-type even after 50 substitutions
(t-test, $p=0.28$).  Similarly, protein length remained fairly 
constant (Figure \ref{f_glen_sel_flat}A), decreasing by $\sim0.4$ aa on average after 50 substitutions. 
We also find that random point mutations were the least deleterious, followed by
insertions, and then deletions (Figure \ref{f_glen_sel_flat}B). In addition, we find that about 70\% of
substitutions were contingent on earlier substitutions. Moreover, about $71\%$ of substitutions were entrenched by 
subsequent substitutions, and $60\%$ were
increasingly entrenched. 
The distribution of forms of epistasis under the alternative landscape was also similar, 
except that there was slightly less negative epistasis and
some substitutions showed no epistasis, due to the presence of exact neutrality among sequences with DOPE scores greater 
than the wild type (Figure
\ref{f_nat_epst_flat}).
Overall, assumptions on the fitness effects of
over-stabilizing mutations have little effect on the dynamics of protein evolution
under purifying selection, starting from the wild-type sequence, likely because
mutations that stabilize are significantly more rare than those that destabilize
\cite{Zeldovich:2007hv}.

\section{Discussion}

We have developed a mechanistic framework for studying the evolution of protein
sequences under purifying selection for native structure and stability. Using the
ligand-binding protein \emph{argT} as a representative example, our results reveal
extensive epistasis between substitutions.  These results suggest a coherent
picture of the role of epistasis in protein evolution under long-term purifying
selection.

We find that while most mutations are nearly-neutral at the time that they fix,
these same mutations would on average have been highly deleterious in earlier
genetic backgrounds. Thus, the substitutions along an evolutionary trajectory are typically 
contingent on epistatic interactions with earlier substitutions. In fact, we find
that a sizable fraction of substitutions are contingent upon the presence of the
immediately preceding substitution.

We also find that after a mutation becomes fixed in a protein, the average fitness
effect of reverting that mutation becomes much more deleterious than reverting it
immediately after it fixed. That is, once a mutation fixes it becomes entrenched
and difficult to remove due to epistatic interactions with subsequent
substitutions. In addition, we observe a modest
tendency for the degree of entrenchment to increase over time. We hypothesize that
these two phenomena---contingency and entrenchment---are ubiquitous in protein
evolution. This is because we expect both phenomena to occur generically in any
fitness landscape that combines the conditional neutrality of mutations with a mode of
protein evolution in which substitutions fix sequentially. In other words,
these phenomena are both consequence of the simple fact that the fitness effects
of a substitution depend on the substitutions that precede it.

\begin{figure}[ht]
	        \centering
	\begin{subfigure}
		\centering
		\includegraphics[width=.45\textwidth]{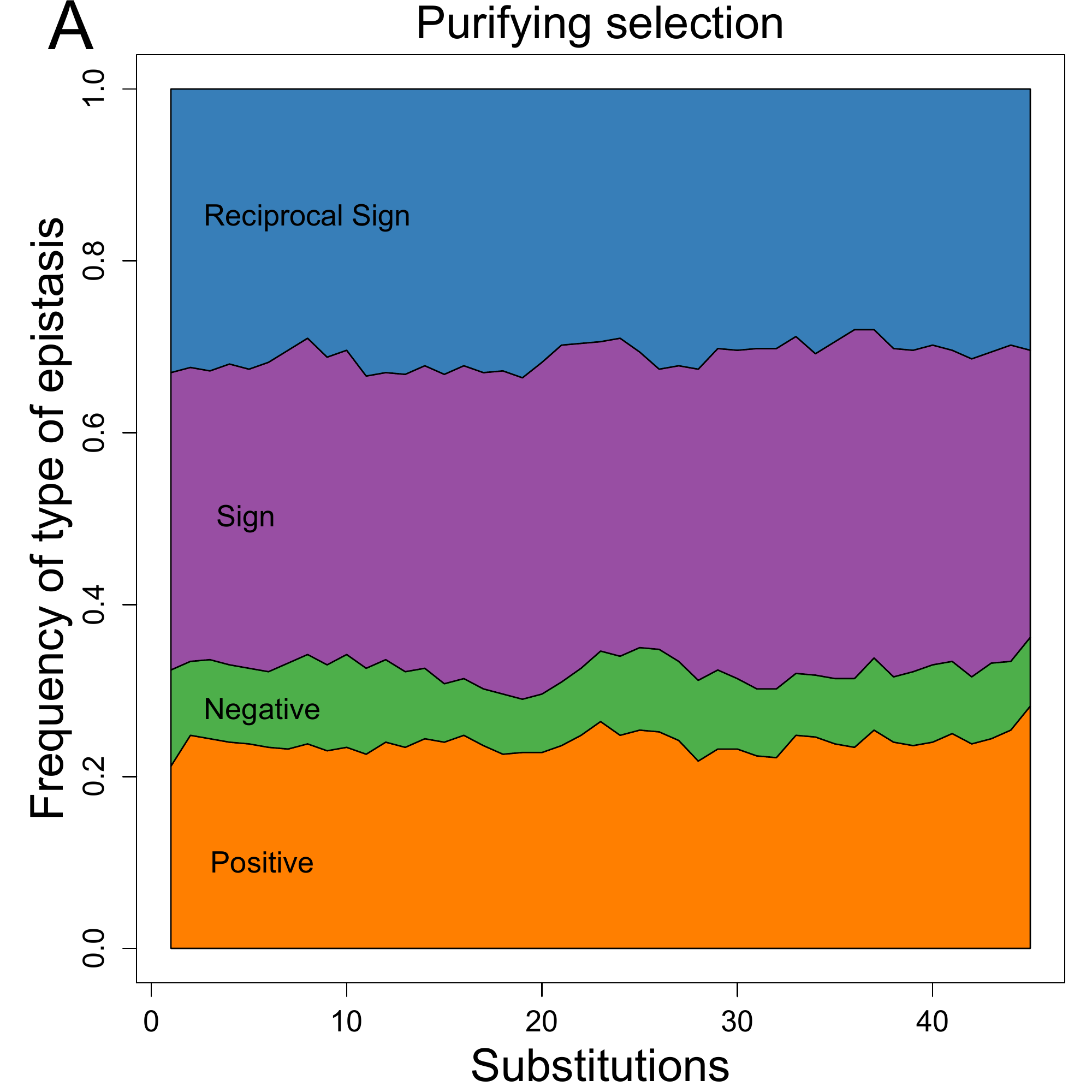}
	\end{subfigure}
	\begin{subfigure}
		\centering
	        \includegraphics[width=.45\textwidth]{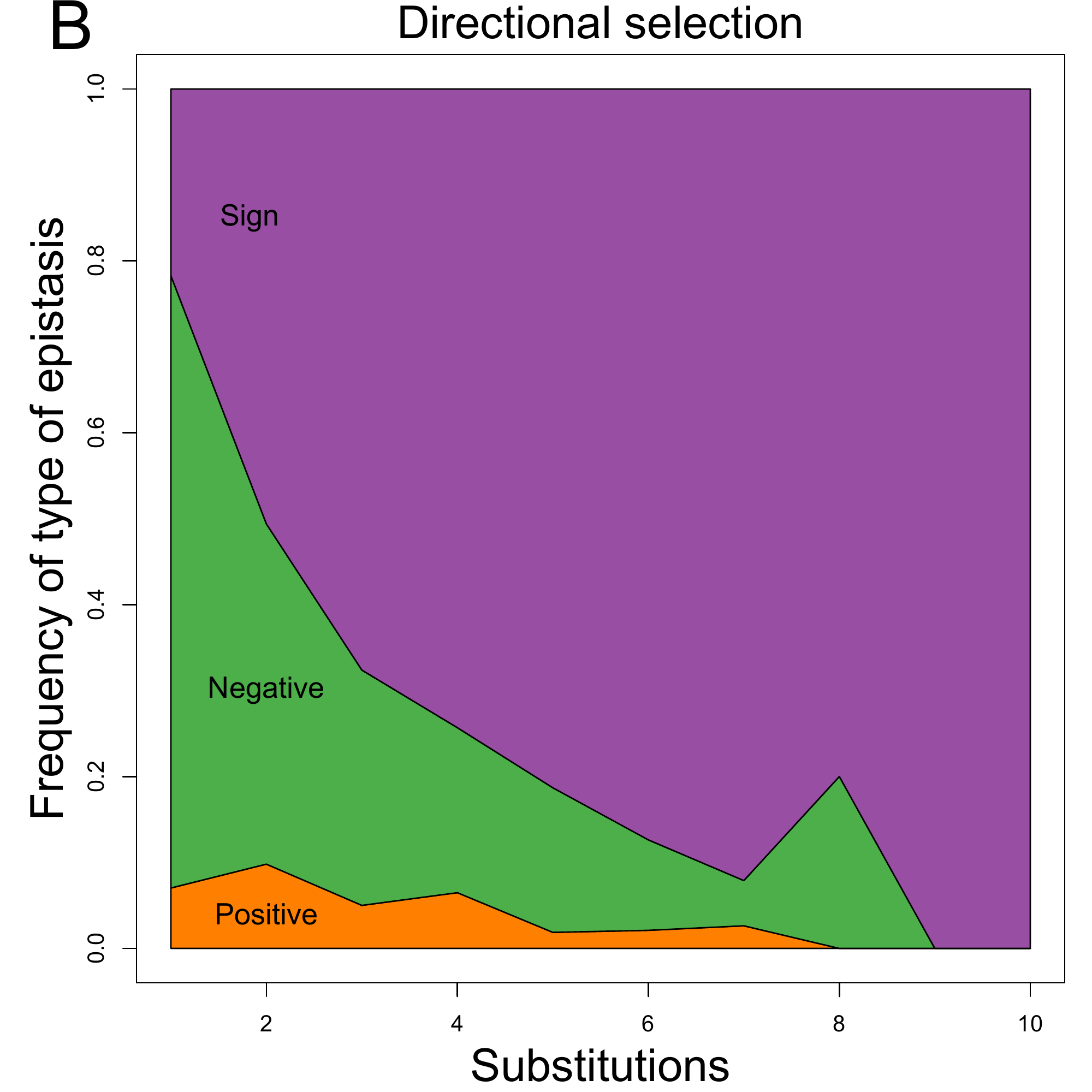}
	\end{subfigure}
	\caption{Forms of epistasis under purifying and directional selection. ({\bf
	A}) Consecutive substitutions that accrue under purifying selection
	tend to be dominated by sign and reciprocal sign epistasis, suggesting that 
the
	geometry of the fitness landscape around the wild-type \emph{argT} sequence 
is
	very rugged. The frequencies of the four forms of epistasis do no change
	substantially along the evolutionary trajectory under purifying selection. ({\bf 
B}) By contrast, under
	direction selection for a new target protein stability, the frequency of the
	four different forms of epistasis
	between consecutive substitutions changes systematically along the
	evolutionary trajectory.
	When
	populations are far from the optimum, negative epistasis dominates; but as
	populations move closer to the optimum, sign epistasis becomes dominant,
	consistent with prior studies of epistasis along adaptive walks \cite{Draghi:2013kg,Greene:2014db}.}
	\label{f_epist_type}	
\end{figure}

Our results also provide insight into several other facets of protein evolution. 
One interesting observation is that the pattern of epistasis under purifying
selection is markedly different than for adaptive substitutions accrued
under directional selection (Figure \ref{f_epist_type}). The nature of epistasis in our
simulations of adaptive evolution agree with other studies of adaptation, which
report that beneficial substitutions tend to be negatively epistatic with each
other \cite{Chou:2011dk,Khan:2011ik,Draghi:2013kg,Greene:2014db}.  
As a result, and in contrast to our results under purifying selection, it
becomes easier to revert an adaptive substitution at later time points in an
adaptive evolutionary trajectory.  Understanding the differences in the nature of
adaptive and nearly-neutral substitutions may provide important insights for
inferring selective regimes from evolutionary trajectories.

By incorporating insertions and deletions into our modeling we have also been able
to address basic questions about the evolution of protein sequence length.  Such
questions are especially relevant because large comparative analyses have shown
that, although homologs in eukaryotes are typically longer than their counterparts
in archaea and bacteria \cite{Brocchieri:2005im,Kurland:2007ho,Wang:2011dp},
protein lengths are highly conserved within eukaryotes and bacteria
\cite{Wang:2005bq,Xu:2006cl}.  Our results confirm these empirical patterns, and
help to explain the mechanistic reason for protein length conservation.  Even
though deletions arise twice as often as insertions, both types of mutations are
observed with equal frequency in homologous sequence comparisons
\cite{delaChaux:2007fg,Chen:2009dc}.  Our analyses confirm this pattern, and
provide a simple explanation: insertions tend to be less disruptive for stability
than deletions, and thus they have higher rates of fixation per mutation.  Our
mechanistic modeling also confirms the long-held view that indels are more
deleterious, as a whole, than point mutations \cite{Taylor:2004bo,Wolf:2007cy,Chen:2009dc,TothPetroczy:2013dl}.

Our results are closely related to several recent studies on protein evolution under purifying selection.  Our analysis of entrenchment confirms the
findings of a recent comparative study by Naumenko et al.~\cite{Naumenko:2012bt} and a simulation study by Pollock et.~al \cite{Pollock:2012ge}, who
described a similar trend as a ``Stokes shift''. We have extended their results to include the entrenchment of insertions and deletions as well.
Moreover, we complement this forward-time perspective by identifying a similar prevalence of contingency under purifying selection, from a
backward-looking perspective. 
More generally, Breen et al.~\cite{Breen:2012fd} have recently argued that epistasis of the form described here---where some substitutions are only
permissible due to preceding substitutions---is the primary factor in molecular evolution. While the formal validity of their inference has been the
topic of debate~\cite{McCandlish:2013fa}, our results here confirm their basic contention and provide a detailed view of the form of epistasis in
proteins under purifying selection.  Likewise, by comprehensively sampling the local fitness landscape around a wild-type protein, McLaughlin et.~al
\cite{McLaughlin:2012hw} and Jacquier et.~al \cite{Jacquier:2013fx} have shown that several stabilizing mutations become destabilizing in the presence
of other mutations, consistent with our observation of extensive sign epistasis between substitution in proteins under purifying selection.  Our results also reflect the notion of
 Wylie \& Shakhnovich that epistatic interactions are governed by underlying bio-physical interactions
between substitutions \cite{Wylie:2011io}.

All of our analysis has been enabled by formulating a mechanistic model that assigns fitness effects to mutations, based on a standard biophysical
description of protein stability.  By contrast, most models of protein sequence evolution along a phylogeny assume independence of sites
\cite{Goldman:1994wf,Kosiol:2007gf,Yang:2008dd,Rodrigue:2010by,Tamuri:2012hv} out of mathematical and computational convenience. Such phylogenetic
models necessarily disregard any possible epistatic interactions between sites.  Although convenient for reconstructing phylogenies or calculating
simple summary statistics, such as $dN/dS$, we know that proteins are in fact highly coordinated structures whose residues often experience
physiochemical interactions that fundamentally determine fold, stability, and function. Our results suggest that incorporating these biophysical
factors, and the resulting non-independence between sites, will be essential for developing accurate models of protein
evolution~\cite{Rodrigue:2010dt,Thorne:2012vk,Wilke:2012df}.
 
The approach we have used here nonetheless makes a number of simplifying assumptions. In particular our evolutionary simulations do not allow
co-segregating mutations -- that is, we assume weak mutation.  As a result, many substitutions along an evolutionary trajectory decrease fitness; and
most of these decreases are compensated by subsequent substitutions.  Although reasonable over evolutionary timescales, relaxing the assumption of
weak mutation would be appropriate for some experimental studies of evolution, and it may alter the dynamics of evolutionary trajectories.

We have also assumed that purifying selection acts on the global stability of a
protein. In reality, however, it is likely that the strength selection on
stability varies within a protein -- so that the protein core experiences stronger
purifying selection than the periphery
\cite{Mirny:1999hh,Koshi:1995jo,Bloom:2006eh}.  Incorporating local stability
requirements would certainly improve our understanding of selective constraints,
but it seems unlikely to qualitatively change our results on the dominant forms of
epistasis that modulate substitutions.

Our analysis has neglected other aspects of purifying selection that likely
operate on proteins -- in particular, selection against adopting alternative
structures \cite{Grahnen:2011hs,Liberles:2012ir}.  Ideally, one could incorporate
negative selection against alternative structures by threading sequences against a
large ``dummy" database of alternative structures.  Such an approach is
computationally prohibitive using homology modeling, at present, when studying an ensemble of evolutionary
trajectories. Whenever in can be feasibly incorporated, however, this additional
constraint may yield important insights into the action of selection as a protein
sequence moves away from the wild-type; as well as insights into the origins of
novel protein folds.

Finally, selection for stability is not the only source of selection on a protein.
A ligand-binging protein, such as considered here, also experiences selection for
its function -- namely, binding its target. Substitutions that are nearly-neutral
with respect to stability might significantly alter the function and will be
unlikely to fix or vice versa \cite{Bloom:2006ch}. However, the number of residues
directly involved in a ligand-binding protein's function is typically small in
comparison to those that predominantly influence its stability
\cite{Bloom:2006ch}. Hence our conclusions regarding epistatic nature of
substitutions are unlikely to be altered substantially by incorporating
constraints on ligand-binding function.

Our approach is fundamentally limited by the accuracy of homology modeling and
stability calculations of mutant protein sequences.  Although homology models
coupled with molecular-dynamics simulations, used here, likely provide greater
accuracy than models based on lattice structures and contact potentials
\cite{Bloom:2006ch,Tokuriki:2007ij,Pollock:2012ge,Ashenberg:2013gd}, they are also
restrictive due to computational cost, and they have required us to sample a
relatively small subset of proposed mutations.  (We have, at least, shown that are
results remain unchanged by increasing sample size 20-fold.) Furthermore, the
accuracy of homology models is reduced as protein sequences diverge from the
wild-type sequence.  As a result, exploring protein sequence evolution beyond a
limited number of substitutions is an imperfect science.  Nevertheless, we find
similar results for the patterns of epistasis between substitutions near the
wild-type sequence as we find between substitutions that occur towards the end of
simulated evolutionary trajectories, which suggests that there is no systematic
bias introduced by decaying accuracy of homology models. Finally, the overall
concordance between our simulation results and comparative sequence analysis, for
questions such as the evolution of protein length, suggests that important and
realistic insights into protein evolution can be derived from the combination of
computational models for protein structures with population-genetic models for
evolutionary dynamics.


\bibliographystyle{apalike}

\clearpage
\pagebreak
	
\clearpage
\pagebreak

\clearpage
\pagebreak

\clearpage
\pagebreak

\clearpage
\pagebreak

\clearpage
\pagebreak

\clearpage
\pagebreak

\setcounter{figure}{0} 
\renewcommand{\thefigure}{S\arabic{figure}}

\begin{figure}[ht]
	        \includegraphics[width=.45\textwidth]{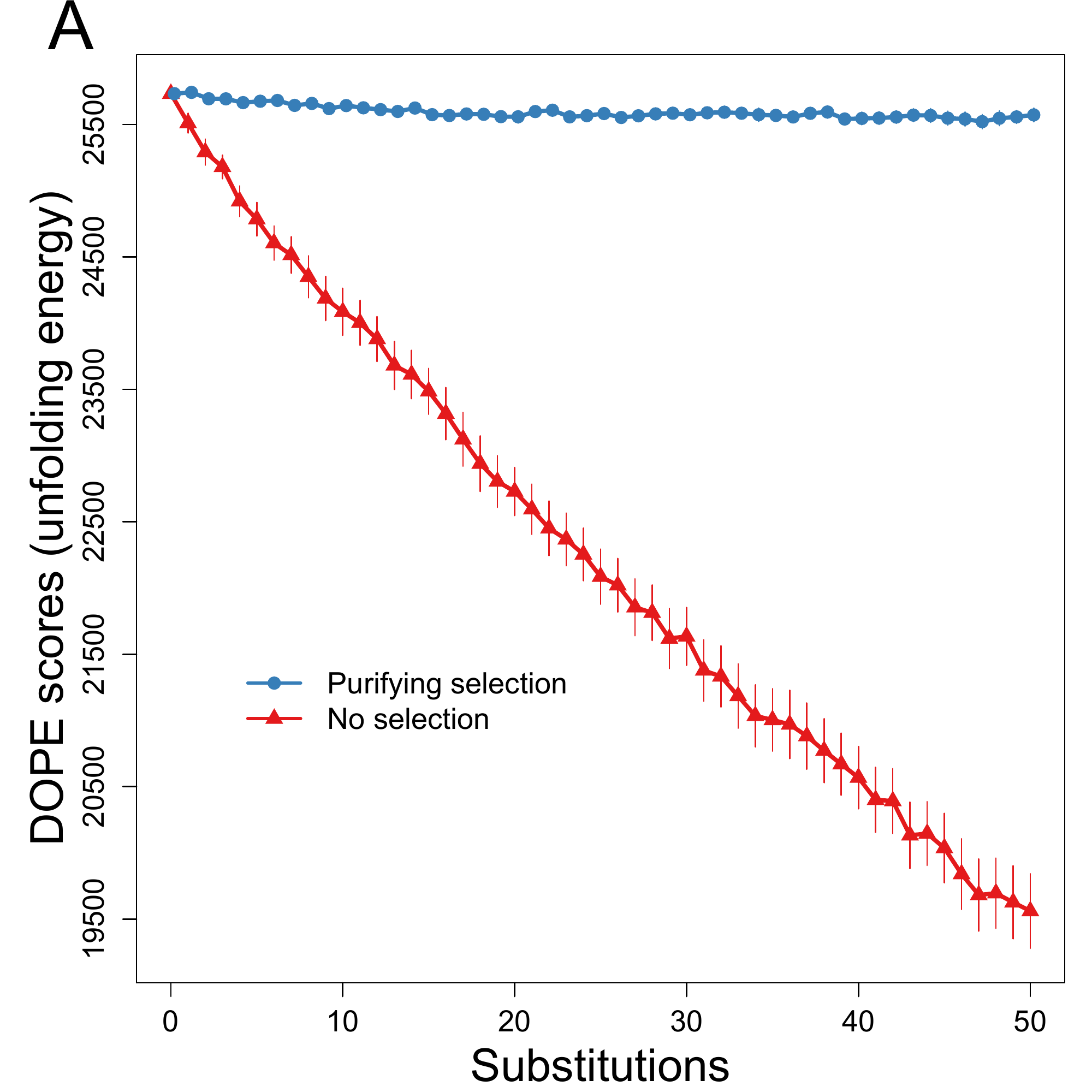}
	\caption{Under
	purifying selection most substitutions are nearly-neutral and do not change 
the stability
	of the protein; whereas in the absence of selection the fixation of random 
insertions,
	deletions and point mutations tends to decrease stability at fitness. 
	Vertical bars indicate $\pm2$ SE around the ensemble mean of
	100 replicate simulated populations.}
	\label{f_dope}	
\end{figure}
\clearpage
\pagebreak

\begin{figure}[ht]
        \includegraphics[width=0.6\textwidth]{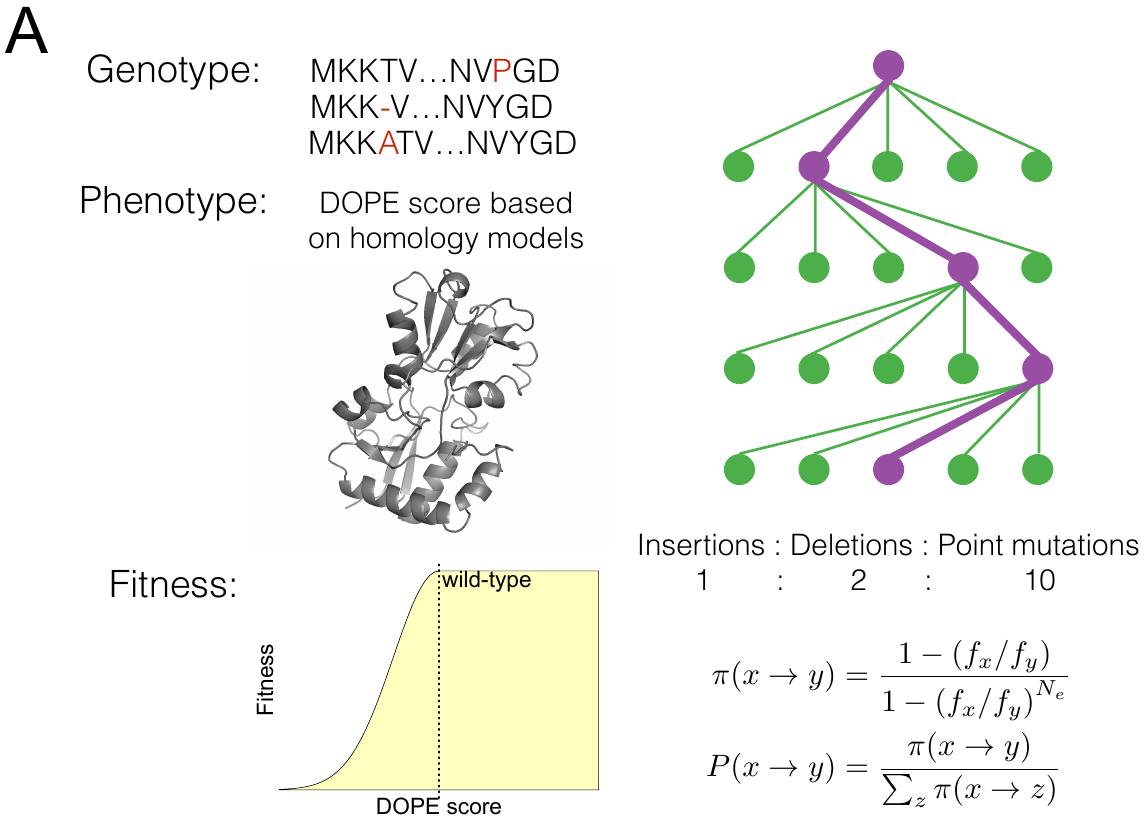}
	\caption{({\bf A}) A model of protein evolution under
	weak mutation and purifying selection against destabilizing mutations. This is 
analogous to
	Figure \ref{f_concept}B, except for using a semi-Gaussian fitness function. Starting from
	the wild-type sequence of \emph{argT} we propose random one-amino acid 
insertions,
	deletions, and point mutations with relative frequencies $1:2:10$. For each of
	the proposed mutants we compute its predicted stability (DOPE score)
	based on homology models, and its associated fitness relative to the stability
	of the wild-type.
	The population fixes one of the proposed mutants, based on its
	relative fixation probability under the Moran model with a population of size
	$N_e$. This process is iterated for 50
	substitutions, which form an evolutionary trajectory. 
	}
	\label{f_alt_model}	
\end{figure}
\clearpage
\pagebreak

\begin{figure}[ht]
        \centering
        \includegraphics[width=.459\textwidth]{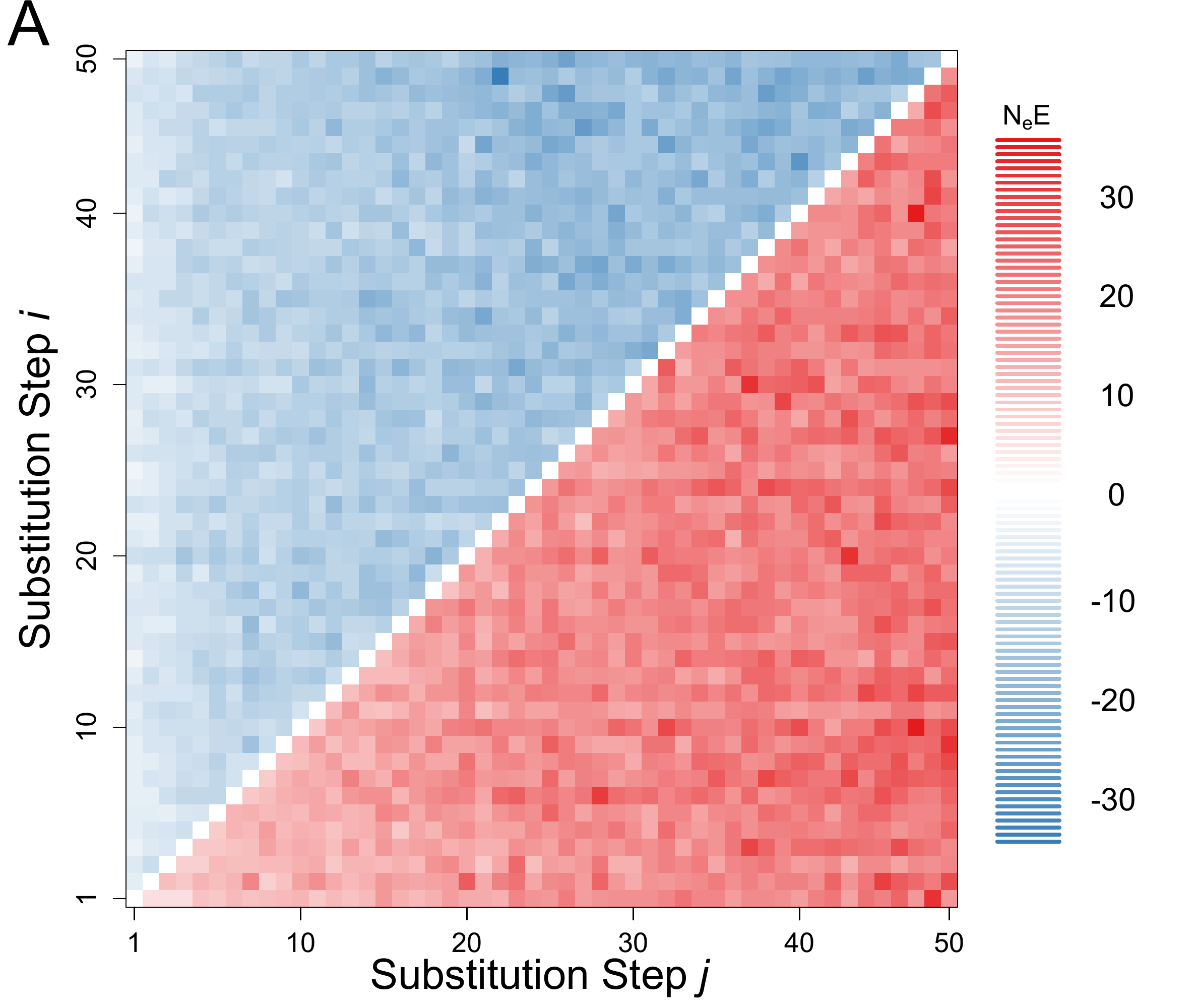}
	\caption{The mean epistatic effect ($E_{(i,j)}$) of focal substitution $i$ at substitution $j$. Each point
	in the grid represents the ensemble average of $E_{(i,j)}$ across 100 replicates. For all $i\ne j$, mean $E_{(i,j)}$
	is significantly different from 0 (t-test, $p<0.02$ for each point in the grid).
}

	\label{f_epist_mat}	
\end{figure}
\clearpage
\pagebreak

\begin{figure}[ht]
	        \centering
	\begin{subfigure}
		\centering
		\includegraphics[width=.45\textwidth]{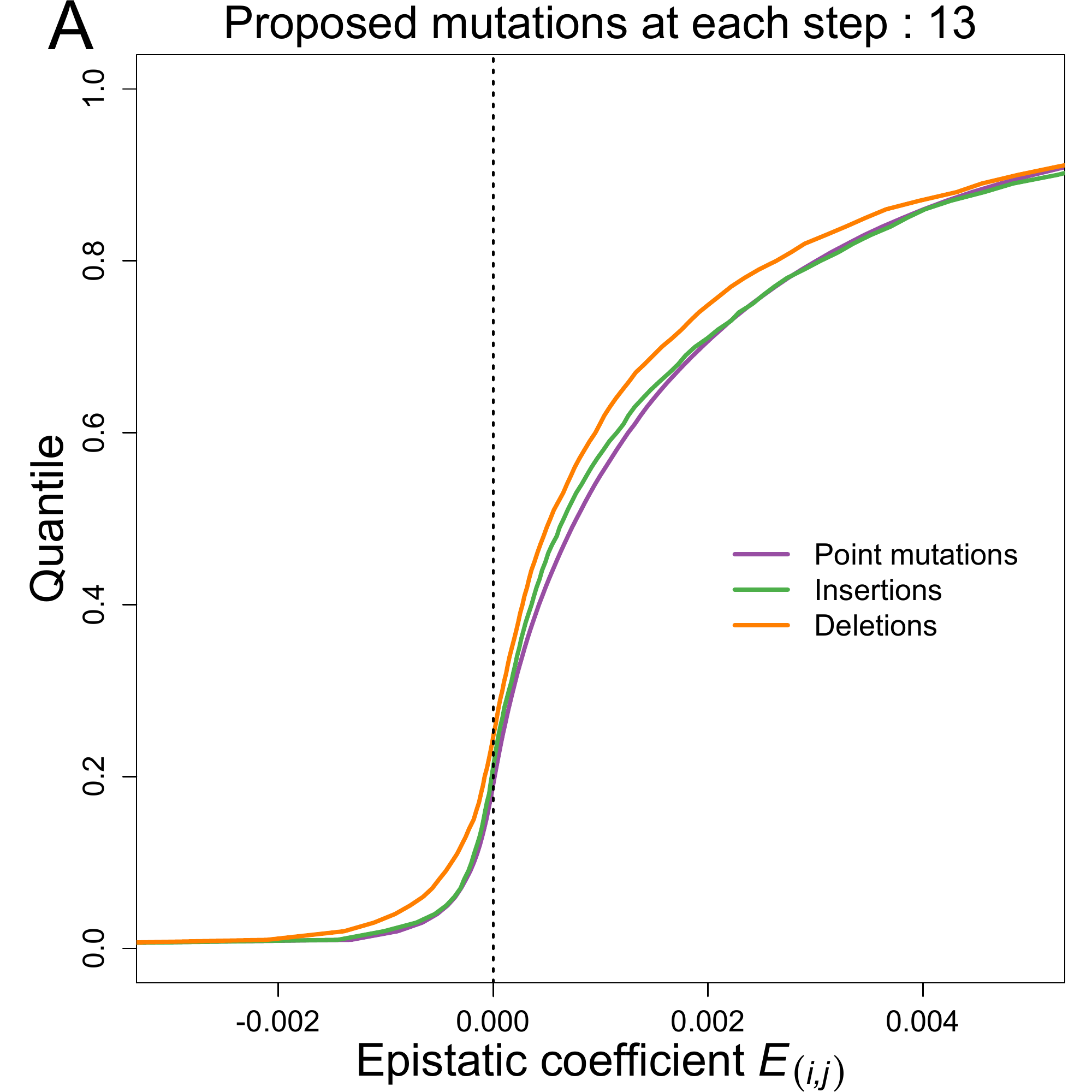}
	\end{subfigure}
	\begin{subfigure}
		\centering
	        \includegraphics[width=.45\textwidth]{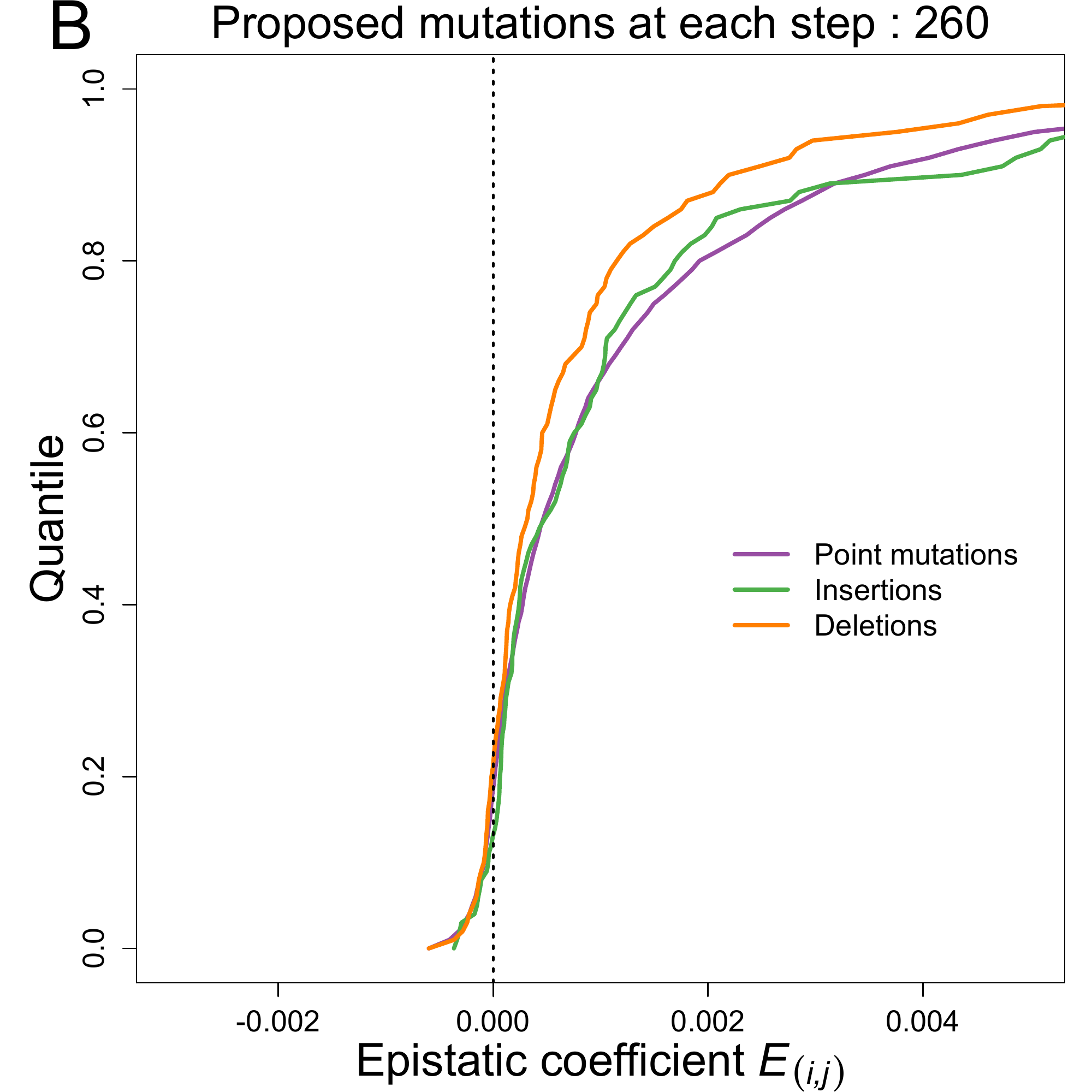}
	\end{subfigure}
	\caption{The distribution of epistatic coefficient ($E_{(i,j)}$) between pairs of 
substitutions along evolutionary trajectories under both low ({\bf A}) and high 
({\bf B})
	numbers of proposed mutations at each step. The epistasis coefficient 
$E_{(i,j)}$ quantifies the effect of reverting a focal substitution $i$ (either
	a point mutation, insertion, or deletion as indicated in the legend) from a 
subsequent genetic background containing $j>i$ substitutions. 
	The distribution of epistatic coefficients are fairly similar among all three 
mutation types. In all three cases over $75\%$ of
	substitutions become entrenched by subsequent substitutions ($E_{(i,j)}>0$), 
under both low and high numbers of proposed mutations at each step.}
	\label{f_epist_low_high}	
\end{figure}
\clearpage
\pagebreak

\begin{figure}[ht]
		\includegraphics[width=.45\textwidth]
{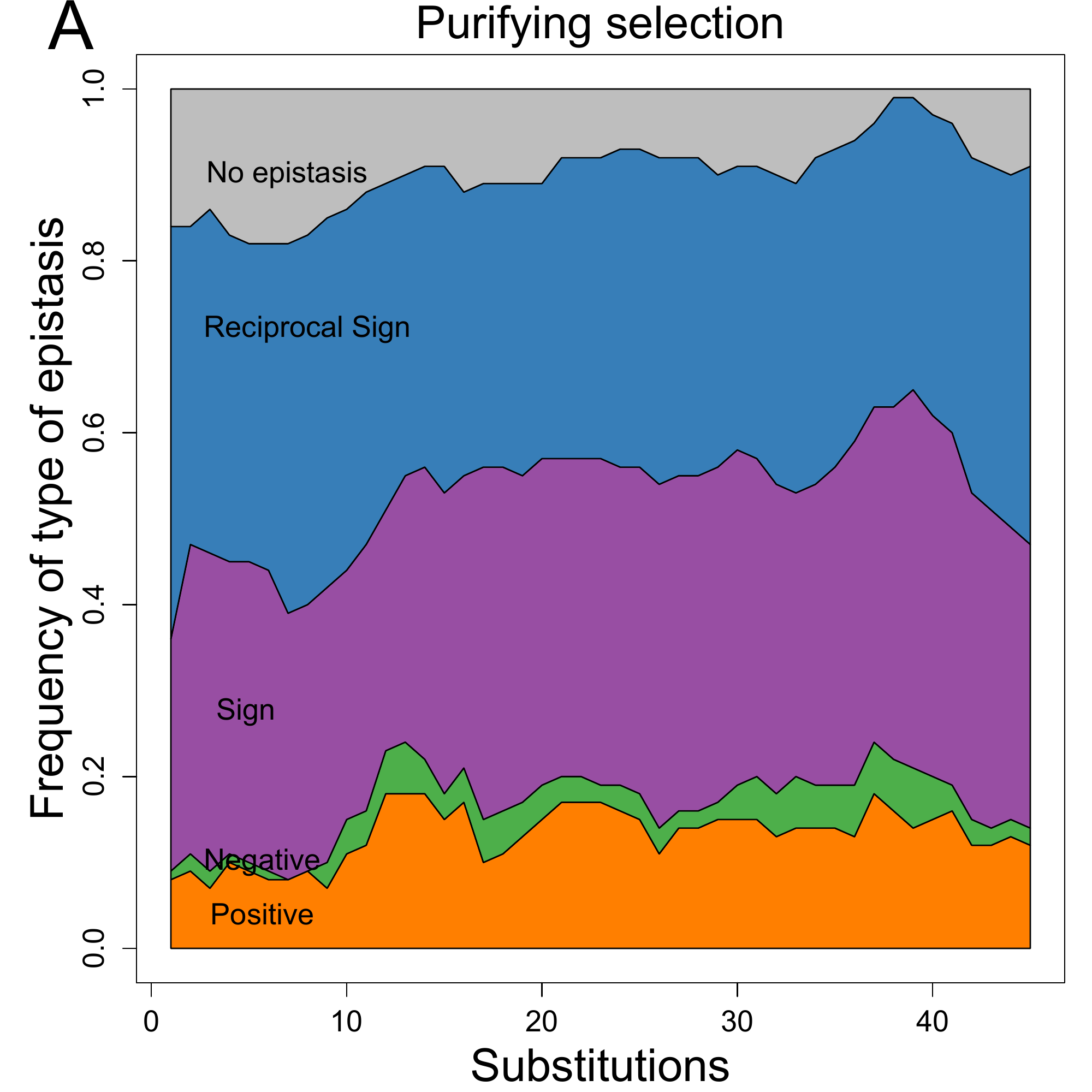}
	\caption{Forms of epistasis under purifying selection against destabilizing 
mutations (semi-Gaussian fitness function). ({\bf
	A}) Consecutive substitutions that accrue under purifying selection
	tend to be dominated by sign and reciprocal sign epistasis. The frequencies 
of the four forms of epistasis do no chasnge
	substantially along the evolutionary trajectory under purifying selection. In 
contrast to a Gaussian fitness function, under
	purifying selection against destabilizing mutations (semi-Gaussian fitness 
function), about $\sim8\%$ of consecutive substitutions are non-epistatic due to the strict neutrality between sequences with DOPE scores greater than the wild type.}
	\label{f_nat_epst_flat}	
\end{figure}
\clearpage
\pagebreak

\begin{figure}[ht]
	        \centering
	\begin{subfigure}
		\centering
		\includegraphics[width=.45\textwidth]{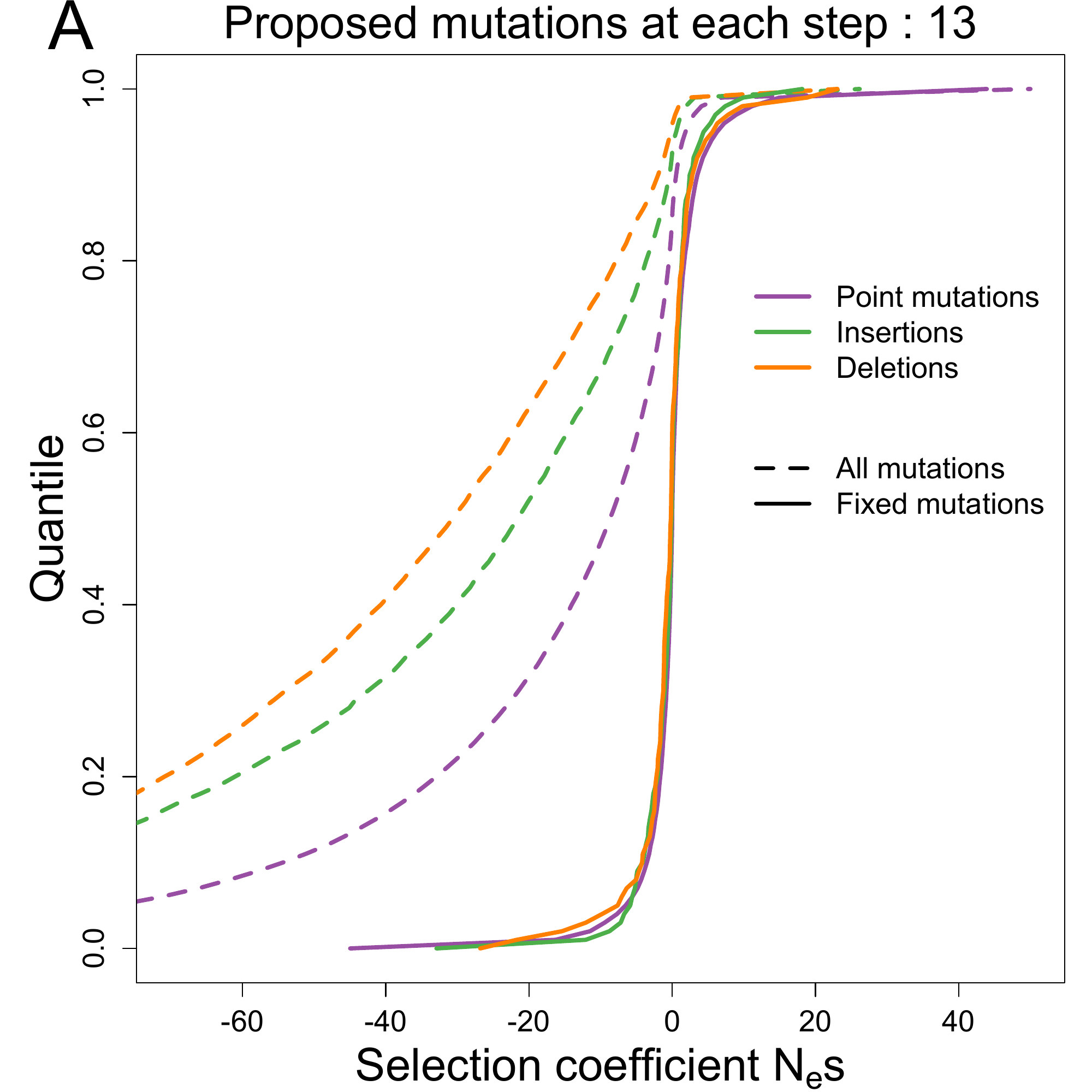}
	\end{subfigure}
	\begin{subfigure}
		\centering
	        \includegraphics[width=.45\textwidth]{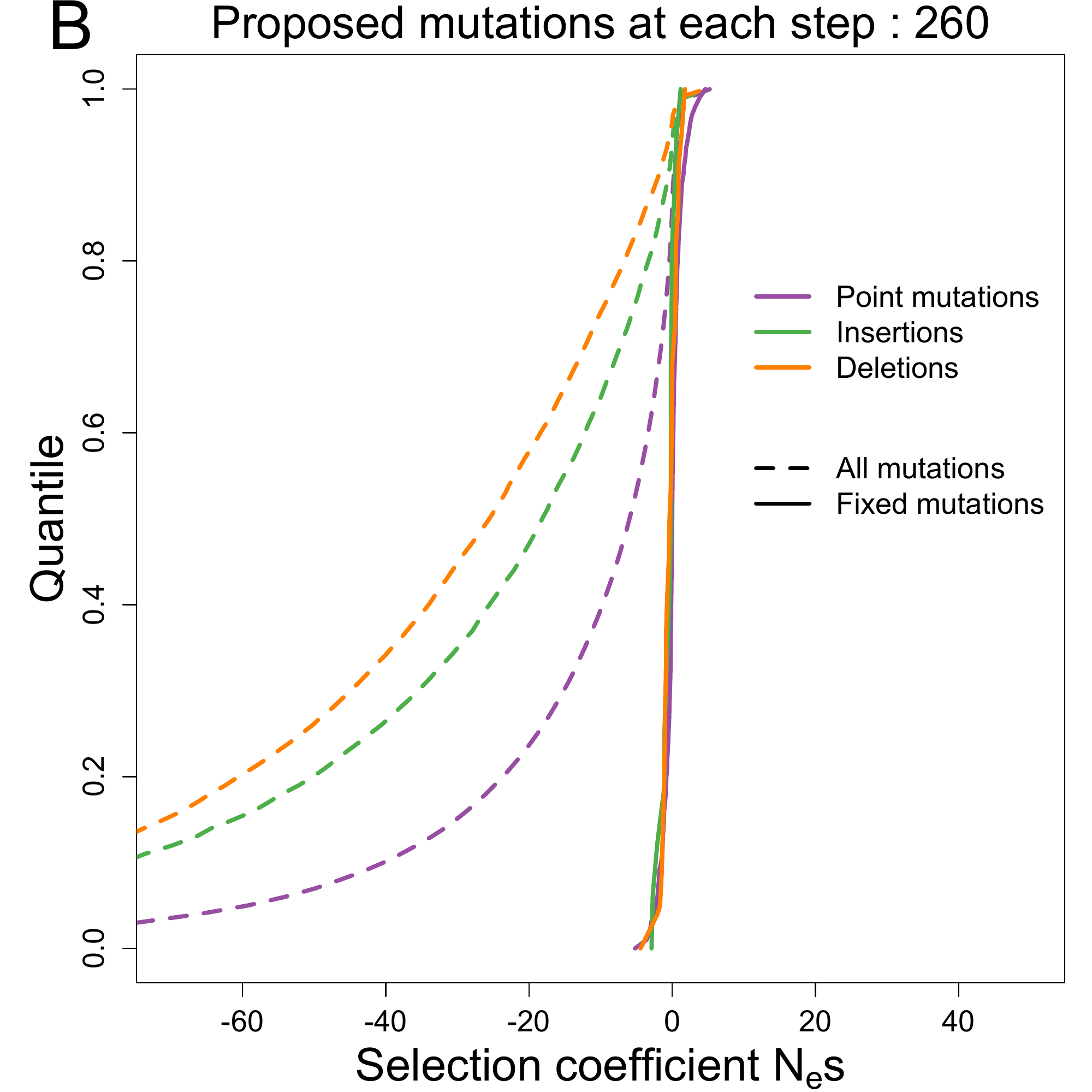}
	\end{subfigure}
	\caption{The distribution of selection coefficients of mutations along 
evolutionary trajectories under both low and high
	numbers of proposed mutations at each step. ({\bf A}) In this panel, identical 
to Figure \ref{f_glen_sel}, we propose 1 insertion, 2 deletions, and 10 point mutations
	at each step. ({\bf B}) In this panel we propose more mutations: 20 insertions, 
40 deletions, and 200 point mutations at each step. Under both scenarios
	we find that random point mutations are the least deleterious, followed by 
insertions and deletions.
	}
	\label{f_sel_low_high}	
\end{figure}
\clearpage
\pagebreak

\begin{figure}[ht]
	        \centering
	\begin{subfigure}
		\centering
		\includegraphics[width=.45\textwidth]{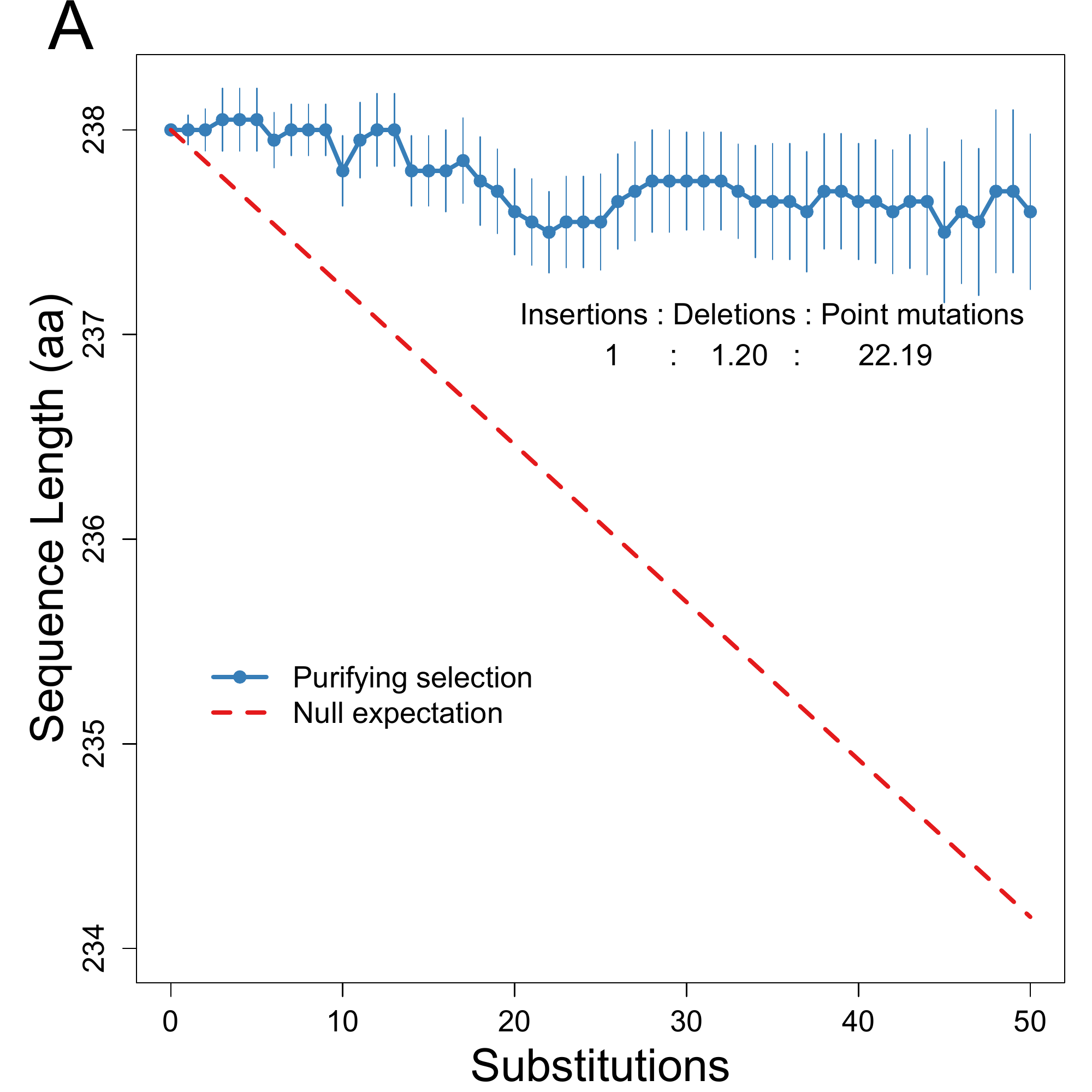}
	\end{subfigure}
	\begin{subfigure}
		\centering
	        \includegraphics[width=.45\textwidth]{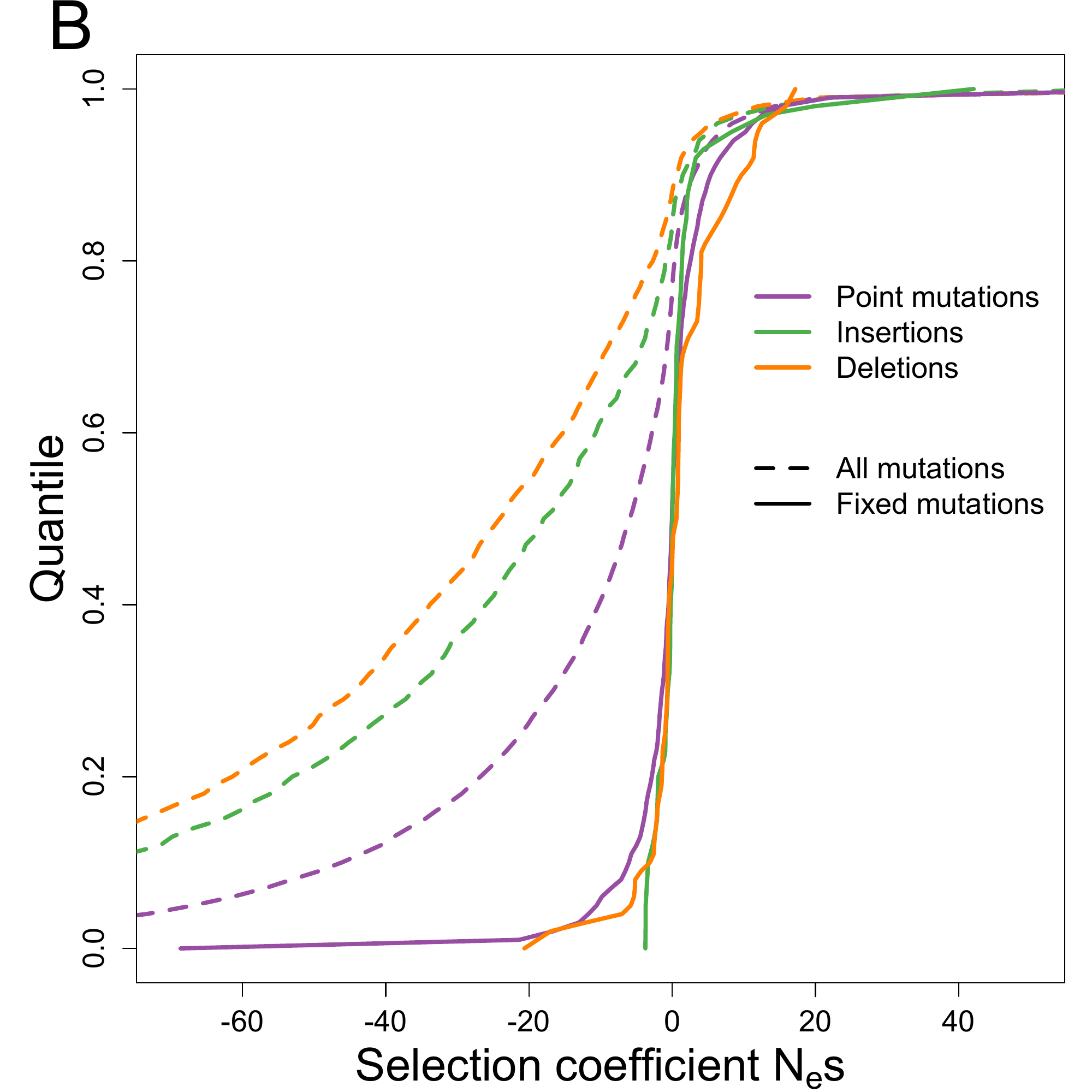}
	\end{subfigure}
	\caption{Protein sequence length remains roughly constant under purifying
	selection against destabilizing mutations (semi-Gaussian fitness function). 
({\bf A}) In the absence of any selection, the sequence length is expected
	to decrease by $\sim3.8$ aa after 50 substitution events
	based on rates of proposed insertions, deletions and point mutations at each
	step. By contrast, under purifying selection  the sequence length decreases 
by
	only $\sim0.5$ aa, on average, after 50 substitutions. Protein length
	is preserved because purifying selection elevates the substitution rate of 
point
	mutations relative to indels; and also decreases the substitution rate of 
deletions
	relative to insertions. Vertical bars indicate $\pm1$ SE around the ensemble 
mean of
	100 replicate simulated populations. ({\bf B}) The distribution of selection 
coefficients
	for random (dotted lines) and fixed (solid lines) insertions, deletions and
	point mutations under purifying selection. Random 
	point mutations are generally less deleterious than random insertions or 
deletions.}
	\label{f_glen_sel_flat}	
\end{figure}

\end{document}